\documentclass[prl,twocolumn,amsmath,amssymb,floatfix]{revtex4}

\usepackage[utf8]{inputenc}

\usepackage{amsmath}
\usepackage{bbm}
\usepackage{graphicx}
\usepackage{bm}

\newcommand{\be}{\begin{equation}}
\newcommand{\ee}{\end{equation}}

\newcommand{\beq}{\begin{eqnarray}}
\newcommand{\eeq}{\end{eqnarray}}

\def\H1{\widehat{H}_1}

\begin{document}

\title{Full counting statistics of photons interacting with emitter}

\author{Mikhail Pletyukhov$^1$, Matou\v{s} Ringel$^2$, Vladimir Gritsev$^2$}
\affiliation{$^1$Institute for Theory of Statistical Physics and JARA -- Fundamentals of Future Information Technology, 
RWTH Aachen, 52056 Aachen, Germany 
\\ 
$^2$Physics Department, University of Fribourg, Chemin du
Musee 3, 1700 Fribourg, Switzerland}

\begin{abstract}
A complete characterization of quantum fluctuations in many-body systems is accessible through the full counting statistics. We present an exact computation of statistical properties of light in a basic model of light-matter interaction: a multimode photonic field coupled to a single two-level emitter. We mostly consider an initial coherent state in a given mode and demonstrate how the original Poissonian statistics gets modified because of quantum many-body scattering effects leading to non-Poissonian distributions. We argue that measuring this statistics in a simple quantum optical setup provides an insight into many-body correlation effects with photons.  
\end{abstract}

\maketitle


{\it Introduction.---}The concept of the Full Counting Statistics (FCS) has been introduced first in the quantum-optical context \cite{Glauber},\cite{Mandel} in order to characterize statistical properties of a non-interacting quantized electromagnetic field. Its fully quantum derivation has been presented in Ref.~\cite{KK}. Later on, this concept has been borrowed and actively developed in the field of mesoscopic physics \cite{LL} for studying statistical properties of electronic currents in meso- and nanoscopic devices for non-interacting and interacting electrons \cite{NB},\cite{BN},\cite{GK},\cite{Schonh},\cite{KS},\cite{BSS},\cite{CBS}. It has been recently shown that the FCS has relationships to a classical-quantum crossover description \cite{SB}, a quantum entanglement \cite{KL}, a characterization of phase transitions \cite{IA}. The FCS of nonlocal observables can be used to quantify correlations \cite{GADP,Hoffer,LF} and a prethermalization behavior in many-body systems \cite{prethermalization}, as well as to define a certain topological order parameter \cite{IA2}. 

Motivated by these developments we revisit the original problem of computing the FCS for photons interacting with an emitter. We  give it the full quantum consideration treating the interaction nonperturbatively.  This study on the interface of quantum optics and nanoscopic physics becomes very actual nowadays, when the two fields are merging together in the continuous progress of fabrication of low-dimensional hybrid photon-solid state nanodevices \cite{QOnano}.  One of the objectives in this interdisciplinary research  is to obtain strong photon nonlinearities as well as a strong photon-emitter interaction for the purposes of an efficient control over individual atoms and phonons. In this respect, the knowledge of statistical properties of an {\it interacting} photon-emitter device becomes essential. 

\begin{figure}[t]
    \includegraphics[width=0.5\textwidth]{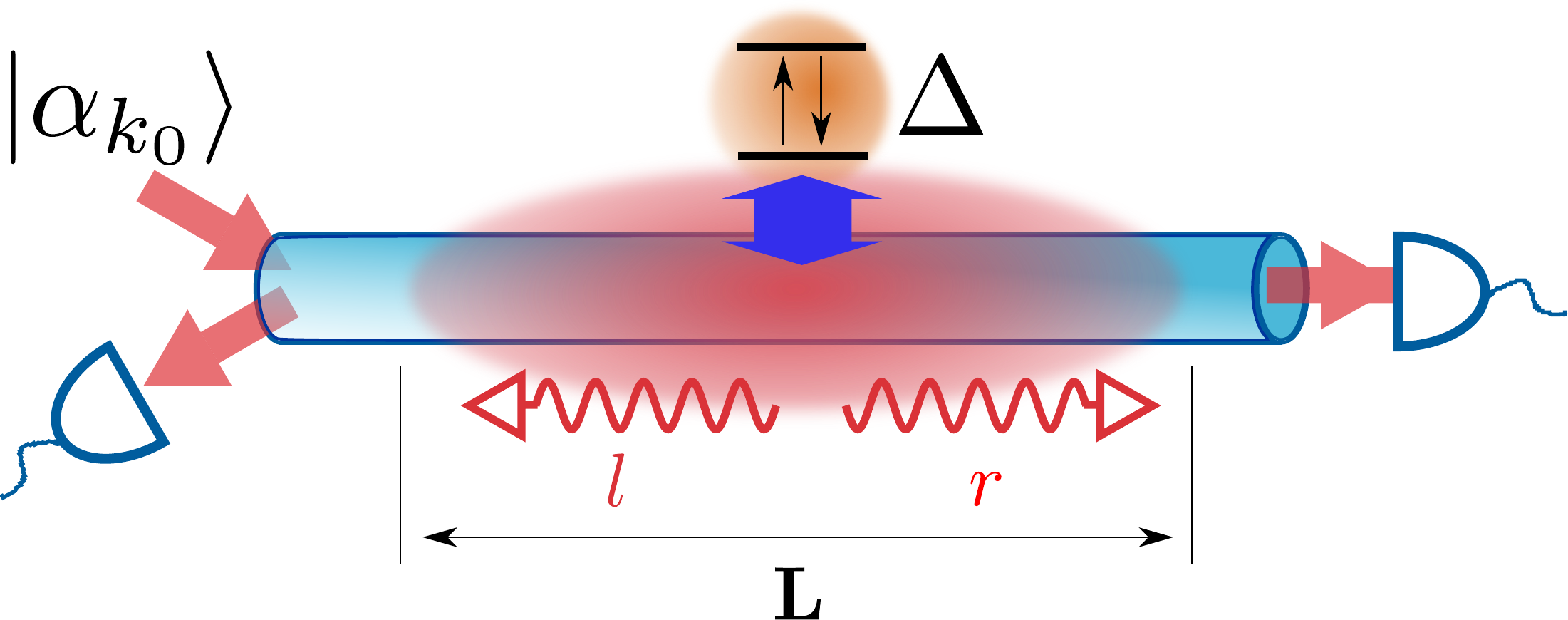}
    \caption{Our system consists of a two-level emitter situated at $x=0$ and coupled to a waveguide (transmission line). A pulse of the length $L$ is prepared in the initial coherent state $|\alpha_{k_0}\rangle$ in the mode $k_0$. We assume the linear spectrum for the radiation field. Because of the interaction with an emitter photons can scatter both forward and backward. The distribution of photons after scattering can be detected by means of existing experimental techniques.}
\label{fig:fcs}
\end{figure}

In this Letter we present an {\it exact} calculation of the FCS in a basic model of light-matter interaction (see Fig.~\ref{fig:fcs}): a multimode propagating photonic field interacting with a two-level emitter. A number of recent studies have focused on a low-dimensional transport properties of a few-photon initial states \cite{few-photons}. Here we significantly generalize and extend these studies in two respects. First we consider rather generic initial states (while mainly focusing on experimentally-easy realizable coherent state). Second, we assume this initial state prepared as a pulse of a size $L$ which introduces an important realistic parameter in photon's statistics. Third, going beyond first and second-order coherences we compute the whole FCS which is a source of all quantum statistical correlations hidden in this fundamental interacting system. Many-body correlations are present here because of a multimode structure of the photonic field interacting with emitter. Interactions considerably  modify the statistics of photons in forward and backward scattering channels in comparison with the Poissonian one of the incident coherent light. In a particular limit of continuous laser field (large $L$) we observe several universal features, e.g. formation of bimodal distribution for arbitrary initial state. We show how tuning the parameters of initial state and pulse duration reveals many-body effects. This can be useful for control of and manipulations with quantum states in emerging nanophotonic devices and circuits. 

{\it The system.---}The most basic model of light-matter interaction is described by the Hamiltonian $H=H_{ph}+H_{em}+H_{ph-em}$, where the free photon part,  $H_{ph}=\sum_{\xi}\int d k \, \xi \, k \, a^{\dag}_{\xi k}a_{\xi k}$, features the linear, unbounded dispersion with two branches $\xi=\pm\equiv r,l$ (right and left); the Hamiltonian of an emitter is approximated by a two-level system with the transition frequency $\Delta$, $H_{em}=\frac{\Delta}{2} \sigma^{z}$; and the photon-emitter interaction is treated in the dipole and the rotating wave approximations, $H_{ph-em}=g_{0}\sum_{\xi}\int d k (a^{\dag}_{\xi k}\sigma^{-}+a_{\xi k}\sigma^{+})$.  As usual, $\sigma^{\pm}$  and $\sigma^z$ denote the Pauli matrices. Note that we consider the one-dimensional waveguide geometry, and therefore the transverse mode index is suppressed.

Introducing symmetric $a_{ek} = (a_{rk} + a_{l, -k})/\sqrt{2}$ and antisymmetric $a_{ok} = (a_{rk}-  a_{l, -k})/\sqrt{2}$ field configurations, we decouple the initial Hamiltonian into the sum  $H=H_{e}+H_{o}$ of  the \emph{even}, 
\beq
H_e = \int dk \left[ k \, a^\dag_{e k} a_{e k} 
+ g    \left( a^\dag_{e k}\sigma^-  + a_{e k} \sigma^- \right) \right] + \frac{\Delta}{2} \sigma_z ,
\label{He}
\eeq
where $g=g_0 \sqrt{2}$, and the \emph{odd}, $H_{o}=\int d k \, k \, a^\dag_{ok} a_{ok}$, contributions. For a local scatterer, one can always choose such a basis in which only one, even, mode is scattered, while the other, odd, mode is decoupled from the interaction Hamiltonian and therefore propagates without scattering. Both even and odd photons are additionally labelled by momentum values $k$ lying on a single branch of the linear dispersion (see the  Supplementary Material \cite{supplement} for details). 

The major consequence of the even-odd decoupling is a factorization of the scattering matrix $S$ into the product $S= S_e S_o$, where $S_o$ equals the identity operator. The matrix $S_e$ corresponding to the model \eqref{He} has been constructed in Ref.~\cite{PGnjp} for all sectors of photon numbers. This construction allows for the full quantum description of scattering of an arbitrary initial field configuration including the coherent light. The scattering of the latter has been also studied in Ref.~\cite{PGnjp}.

In order to completely specify the model presented in Fig.~\ref{fig:fcs}, we define the initial state to be $|\mathrm{in} \rangle =|\alpha\rangle_{r,k_0} \otimes |\downarrow \rangle$, where the incident right-moving photons are prepared in the coherent state $|\alpha \rangle_{r, k_0} \equiv |\alpha \rangle_{r} =  D_{r,k_0} (\alpha) |0 \rangle$, and the two-level emitter is initially in the ground state $|\downarrow \rangle$. Here $|0 \rangle$ denotes the photonic vacuum, and $D_{r,k_0} (\alpha) \equiv D_r (\alpha)= \exp (\alpha b^{\dagger}_{r, k_0} - \alpha^* b_{r, k_0}) $ is the displacement operator. The latter is composed of the operators $b_{r,k_0}^{\dagger}  \equiv b_{r}^{\dagger} = \frac{1}{\sqrt{L}} \int_{-L/2}^{L/2} d x a^{\dagger}_{r} (x) e^{i k_0 x} $ creating wave packets centered around the given mode $k_0$ of the right branch of the original spectrum with the width $2 \pi/L$ (see the Supplementary  Material \cite{supplement} for details). Analogously we define $b_{l,-k_0}^{\dagger}  \equiv b_{l}^{\dagger} = \frac{1}{\sqrt{L}} \int_{-L/2}^{L/2} d x a^{\dagger}_{l} (x) e^{-i k_0 x} $. The length scale $L$ can be associated with the spatial extension of the initial pulse, see Fig.~\ref{fig:fcs}. The mean number of the wave packets in the state $|\alpha \rangle_{r}$ is given by $\bar{N} = |\alpha|^2$.  It is also convenient to introduce the dimensionless coupling strength $\gamma=\pi g^{2}L$ and detuning $\delta=(k_{0}-\Delta)L$. To facilitate computations we represent the initial state in the even-odd basis
\beq
|\alpha \rangle_{r} =D_e \left(\frac{\alpha}{\sqrt{2}}\right)D_{o}\left(\frac{\alpha}{\sqrt{2}}\right)|0\rangle,
\label{factorize}
\eeq
where the displacement operators $D_{e,o}$ are defined with help of mutually commuting operators $b_{e, k_{0}}^{\dagger}$ and $b_{o,k_{0}}^{\dagger}$, respectively. 

The statistics of the initial field, defining a probability $p_{\alpha} (n)$ to find $n$ photons in
the mode $k_0$, is given by the Poissonian distribution $p_{\alpha} (n)= e^{-\bar{N}} \frac{\bar{N}^{n}}{n!}$.
Due to the presence of the photonic dispersion,  photons can leak from the mode $k_0$ to other modes on both branches of the spectrum by
virtue of scattering processes, what induces nontrivial correlations between photons. Statistical properties of the incident beam are thus being changed. A fraction of photons is reflected, and their statistics is also of great interest. We propose a calculation of the FCS in both forward and backward scattering channels, which is {\it exact} and thereby {\it nonperturbative} in both $g$ and $\bar{N}$.

{\it Defining FCS.---}Generally speaking, FCS can be defined as a  generating function $F(\lambda)
=\sum_{n=0}^{\infty} e^{i \lambda n} p (n)$ associated with  a probability  distribution $p(n)$ of
photon numbers $n$ in some specific photonic state. The function $F (\lambda) $ ($\ln F (\lambda)$)
generates $m$-th order moments (cumulants) of the distribution $p (n)$: One has simply to consider the $m$-th derivative with respect to $i \lambda$ at $\lambda=0$. Depending on a physical meaning of the distribution $p (n)$, one can distinguish between different types of FCS.

It is convenient to introduce the variable $z =e^{i \lambda}$ analogous to fugacity. In these terms, the Fourier expansion of the $2\pi$-periodic function $F (\lambda)$ acquires a form of power series in $z$, $F (z) = \sum_{n=0}^{\infty} z^{ n} p(n)$. As a function of $z$,  $F (z)$ can be analytically continued inside a  circle of the unit radius $|z| \leq 1$. In particular, one can treat $z$ as a real-valued parameter on the interval $-1 \leq z \leq 1$. Under this constraint the function $F(z)$ is real-valued. The probability normalization 
$\sum_{n=0}^{\infty} p(n) = F(\lambda=0) =F(z=1)  = 1$
follows from the normalization of a quantum state whose statistical properties are being studied. 

In the FCS framework we want to characterize the scattering state $| \mathrm{out} \rangle = S | \mathrm{in} \rangle$. Due to the factorization property \eqref{factorize} this state can be represented as $| \mathrm{out} \rangle = | \mathrm{out} , \alpha/\sqrt{2} \rangle_e \otimes | \alpha/\sqrt{2} \rangle_o$, where $| \mathrm{out} , \alpha/\sqrt{2} \rangle_e$ is the scattering state in the model \eqref{He} with the single (even) branch of the spectrum \cite{PGnjp,supplement}, and $| \alpha/\sqrt{2} \rangle_o$ is the coherent state in the odd sector. Note that in both cases the coherence parameter equals $\alpha/\sqrt{2}$.

While focusing mainly on the coherent initial state we define the momentum-resolved FCS of photons in the scattering state $|out\rangle$ corresponding to the initial state $|in\rangle\equiv|\psi\rangle$
\beq
F_{\psi} (\lambda_r , \lambda_l) = \langle \mathrm{out} | e^{i \lambda_{r} N_{r}  + i \lambda_l N_l} | {\mathrm{out}} \rangle,
\label{eq:F-def}
\eeq
with help of the  wave packet number operators \cite{supplement}
\be
N_{\xi} = \sum_{n=0}^{\infty} n | n \rangle_{\xi} \,\, _{\xi}\langle n | ,
\quad | n \rangle_{\xi} = \frac{(b_{\xi}^{\dagger})^{n}}{\sqrt{n !}} |0 \rangle,
\quad \xi =r,l
\ee
which possess nonnegative integer eigenvalues. Using the identity  $e^{i \lambda_{\xi} N_{\xi}} = 1+ \sum_{n=1}^{\infty} (z_{\xi}^{n} - 1)
| n \rangle_{\xi} \,\, _{\xi}\langle n|$, where $z_{\xi}=e^{i\lambda_{\xi}}$,
we express \eqref{eq:F-def} for $|\psi\rangle=|\alpha\rangle_{r}$ as
\beq
F_{\alpha} (\lambda_r , \lambda_l)\!\!
&=&\!\!1 +\sum_{n=0}^{\infty} (z_{r}^{n}-1) p_{\alpha} (n) |s_{n0} |^2 \\
&+& \sum_{m=0}^{\infty} (z_{l}^{m}-1) p_{\alpha} (m) |s_{0m} |^2 \nonumber \\
&\!\!\!+&\!\!\!e^{\bar{N}}\!\!\!\sum_{n,m=0}^{\infty} (z_{r}^{n}-1)(z_{l}^{m}-1) p_{\alpha} (n) p_{\alpha} (m)  |s_{nm} |^2\nonumber
\label{eq:DefinitionOfFalpha}
\eeq
where the coefficients $s_{nm}$
quantify how much the FCS \eqref{eq:F-def} differs from the initial, Poissonian one (for which all
$|s_{nm}| =1$). 
The forward and backward scatterings are characterized by coefficients $s_n^r:=s_{n0}$ and
$s_n^l:=s_{0n}$, respectively.
A computation of the coefficients exploits the explicit form of the $|\mathrm{out}\rangle$ state \cite{PGnjp}. As a result, we find \cite{supplement} the following expressions
\beq
s_{nm} &=& \sum_{p=0}^n \binom{n}{p} \left(\frac{-\gamma}{2\rho}\right)^{p+m} c_{p+m}(\rho),
\label{s-nm}
\eeq
where $\rho=(\delta+i\gamma)/2$, $c_{n}(\rho)=\rho e^{i\rho}[j_{n-1}(\rho)-ij_{n}(\rho)]$ and $j_{n}(\rho)$ are the spherical Bessel functions of the first kind. In the limits $\gamma \to 0$ and $\gamma \to \infty$ we obtain $s_n^{r} = 1$, $s_n^{l}=0$ (the whole pulse is transmitted) and $s_n^{r}=0$, $s_n^{l} =(-1)^n$ (the whole pulse is reflected), respectively. In both cases the Poissonian statistics is retained.

\begin{figure}[h]
       \includegraphics[width=0.48\textwidth]{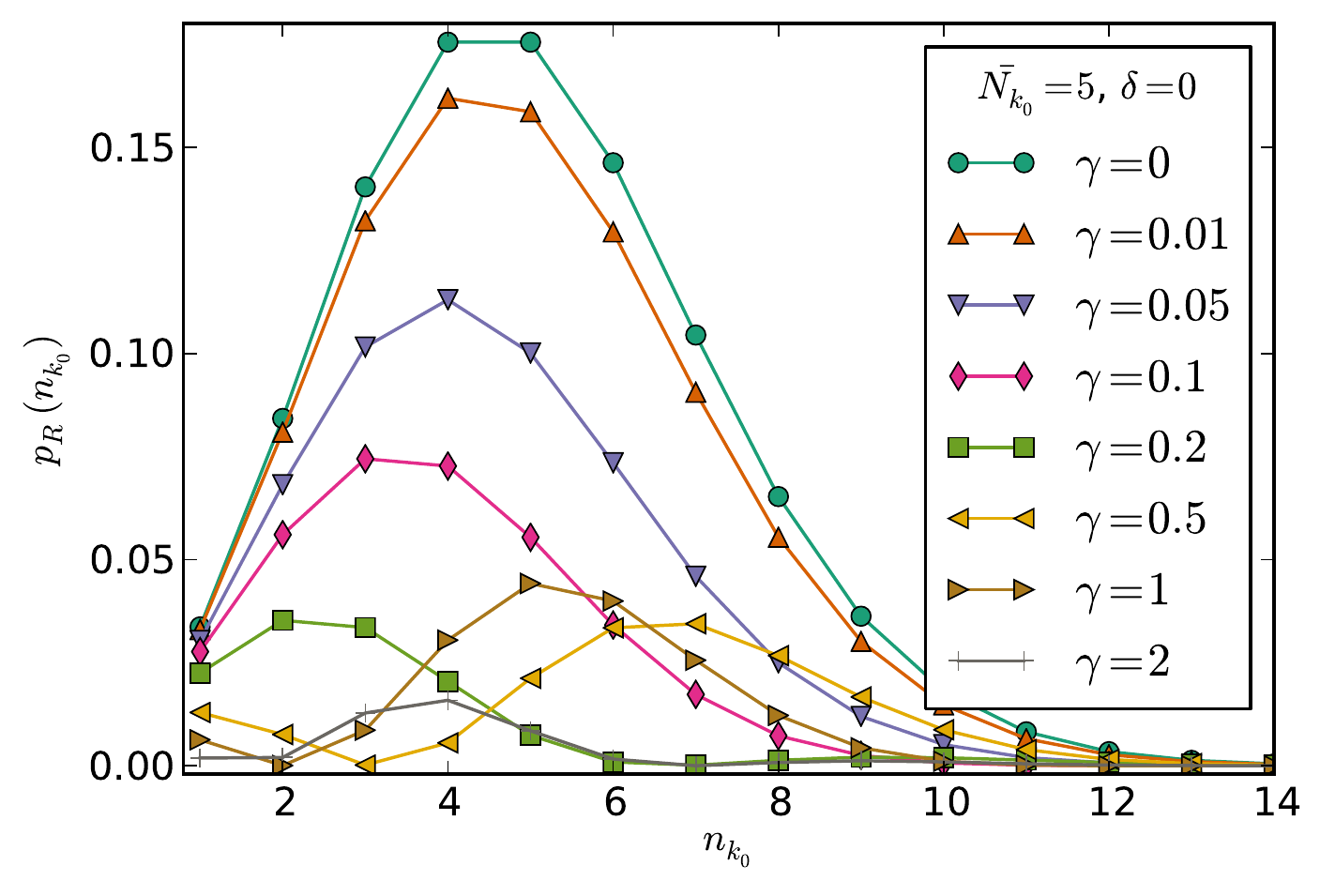}
    \caption{The probability
    distribution~$p_{r}(n)$ for the number of particles in the mode~$k_0$ in the forward-scattering channel, plotted for various values of coherent state parameter $\bar{N}$ and the field-emitter interaction
    strength~$\gamma$. Distributions for a different values of $\bar{N}$ are plotted in the Supplement.
    }
    \label{fig:r-fun}
\end{figure}

\begin{figure}[h]
       \includegraphics[width=0.48\textwidth]{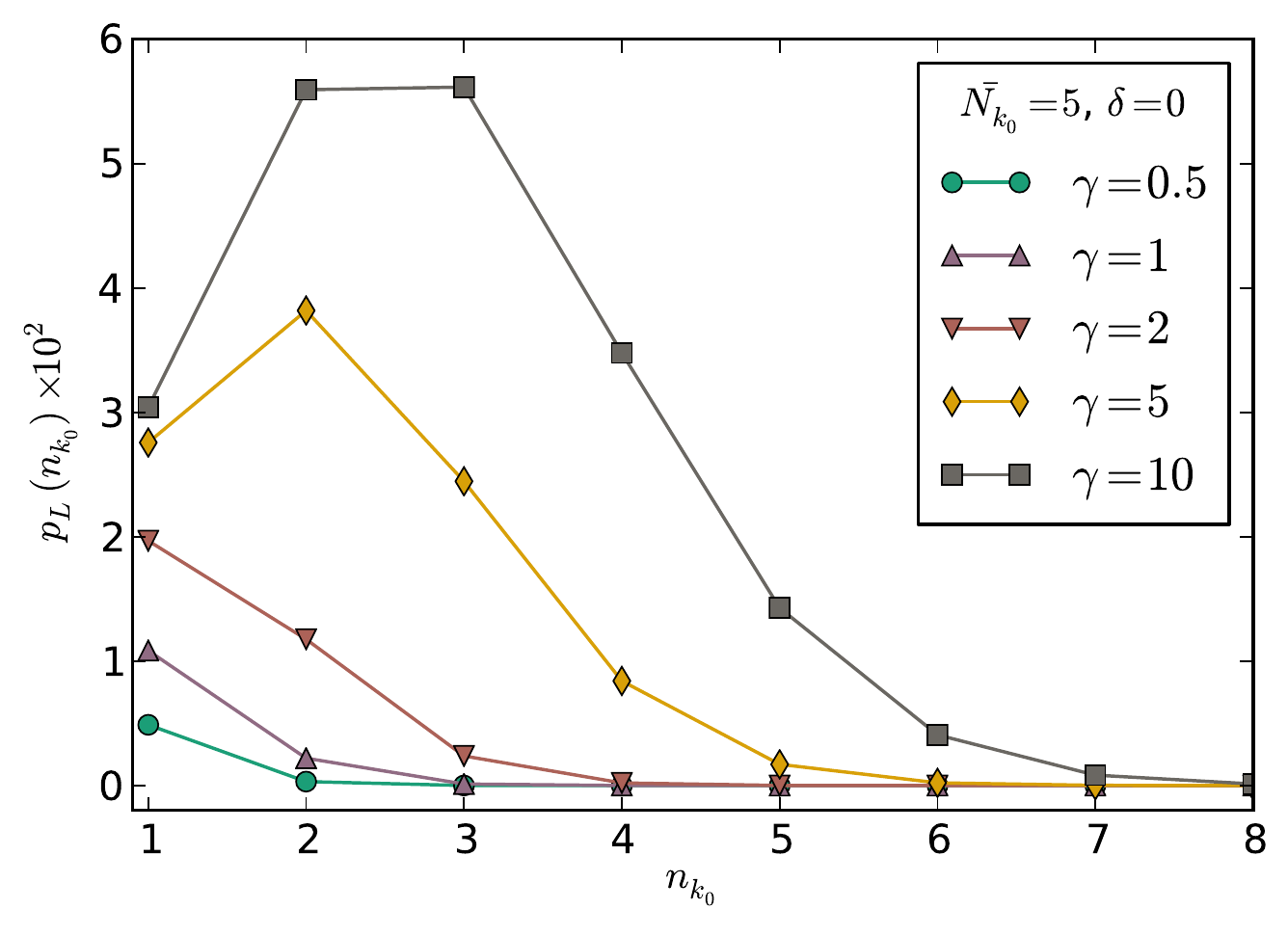}
    \caption{The probability
    distribution~$p_{l}(n)$ for the number of particles in the mode~$k_0$ in the backward-scattered channel, plotted for various values of coherent state parameter $\bar{N}$ and the field-emitter interaction
    strength~$\gamma$. Distributions for a different values of $\bar{N}$ are plotted in the Supplement.
    }
    \label{fig:l-fun}
\end{figure}

{\it Results and discussion.---}
In Figs.~\ref{fig:r-fun} and \ref{fig:l-fun} we plot the probabilities $p_{r,l} (n) = p_{\alpha} (n) |s_n^{r,l}|^2$ associated with \eqref{eq:F-def} which describe the distribution of transmitted/reflected photons in the mode $k_0 \,\, (-k_0)$ after scattering. The mean numbers of photons corresponding to these distributions are presented in the Supplementary Material \cite{supplement}.

Analyzing the FCS we elucidate several important results coming from the many-body character of scattering off the two-level system. First, the resulting distributions of photon numbers becomes essentially non-Poissonian when the interaction is switched on. Second, although the absolute value of density of particles in the mode $k_{0}$ is reduced as a function of $\gamma$ this evolution is very non-trivial: $p_{R}(n_{k_{0}})$ is non-monotonous as a function of $\gamma$ with a tendency towards forming a peak at fewer particles number $n$ while at some $\gamma(\bar{N})$ a new peak emerges at some $n^{*}>\bar{N}$ with a larger height.  Third, the behavior of probabilities is weakly oscillatory as a function of $n$. One can understand these effects as a competition between scattering into other-$k\neq k_{0}$ modes, backscattering and the initial Poisson distribution.

To gain further insight we first analyze the situation of {\it continuous} laser radiation, $L\rightarrow\infty$. This corresponds to $\rho\rightarrow\infty$ with the constant ratio $\delta/\gamma$. In this limit a number of {\it universal} results emerge which can be characterized in terms of transmission $T=|t|^{2}$ and reflection $R=1-T$ probabilities. Here $t=\delta/(\delta+i\gamma)$. Analytical results for FCS (see Supplement for details) $F(\lambda_{r},\lambda_{l})=1+e^{-\bar{N}_{l}}f_{r}+e^{-\bar{N}_{r}}f_{l}+f_{r}f_{l}$ where $f_{\xi}=\exp(\bar{N}_{\xi}(e^{i\lambda_{\xi}}-1))-1$ and $\bar{N}_{\xi=r,l}=\{\bar{N}T,\bar{N}R\}$ suggest that the distribution for transmitted photons  $(\lambda_{l}=0)$ is {\it bimodal}, $p(n)=(1-e^{-\bar{N}_{l}})\delta_{n,0}+e^{-\bar{N}_{l}}p_{\alpha_{r}}(n)$ where $|\alpha_{\xi}|^{2}=\bar{N}_{\xi}$. Similar results hold for reflected photons. 
Moreover, in this limit we are able to compute the FCS for an arbitrary initial state. In particular, for the Fock (number) initial state $|N\rangle$ we again obtain a bimodal distribution $p_{N}(n)=\delta_{n,0}(1-T^{N})+\delta_{n,N}T^{N}$ in transmission  channel. This result is highly non-classical and reflects the quantum nature of the number state: either all $N$ photons are simultaneously transmitted or all reflected. Similar bimodal structure of the distribution takes place for the squeezed state. Explicit result for the $F_{\psi}(\lambda_{r},\lambda_{l})$ for the arbitrary initial state $|\psi\rangle$ is given in Supplement. The FCS in the right/left channels has an interesting factorization structure: the factors $p_{\psi}(n)$ specific for the initial state $|\psi\rangle$ multiplies by the factors $|s_{nm}|^{2}$ coming from the scattering itself. This leads to the bimodal structure of the resulting $p(n)$. 

In the {\it pulsed} laser regime (finite $L$) the factors $s_{n}$ have a quasi-periodic oscillatory (in $n$) behavior which can be approximated as $s_{n}^{r}\sim\cos(\sqrt{\gamma n/2})$ in the limit of large $n$.  For generic $n$ these functions of $\gamma$ and $\delta$ are plotted in the Supplement. From this we infer that when the maximum of the Poisson distribution $p_{\alpha}(n)$ ($\sim \bar{N}$) coincide with one of the maximum of the $|s_{n}^{r}|^{2}$ function we get an enhancement of the scattering in the $n^{*}$-particle channel. This manifestation of the many-body scattering effect is visible in Figs. (\ref{fig:r-fun}) as a re-entrant effect described above. This phenomenon suggests an experimental way to tune to a resonance with $n$-particle scattering. We note however that the available phase space volume for the irreducible $n$-particle scattering gets smaller as $n$ increases (see e.g. \cite{Mussardo} for generic arguments) and therefore many-body effects are suppressed by these natural causes.

Scattering to the other $k$-modes is not directly visible from the FCS we computed here. In a sense, other-then-$k_{0}$ channels act as dissipation with respect to the $k_{0}$ channel. Careful analysis of the $n$-particle $S$-matrix \cite{PGnjp} reveals that a single-particle scattering to the $k\neq k_{0}$ modes is suppressed by the energy conservation while essentially many-body scattering effects coming from the irreducible part of the $n$-particle scattering are generically suppressed as $(k-k_{0})^{-n}$ (this power is stronger at the resonance). This shows that while single-particle scattering to $k\neq k_{0}$ is suppressed by energy conservation, irreducible $n$-particle scattering is suppressed by the power of $n$ and phase-space arguments.

We comment further on the previous studies of statistics in this quantum optical setup. Indeed, after the work of Mollow \cite{Mollow} there were many studies of phenomenon now referred as a resonance fluorescence. Statistics of {\it classical} light was studied long ago in \cite{earlier} in a system of two-level atom coupled to a {\it single}-mode photonic field. In this approach a strong coherent field approximation has been used (which results in replacement of operators by their coherent-state expectation values) as well as a Markovian approximation \cite{BB}. Here we avoid these approximations.  Moreover we can exactly evaluate several other FCS-like functions which clearly show generically non-Markovian character of a scattering in our setup. This studies will be reported elsewhere \cite{PRG-long}.

{\it Experimental scheme}.
To measure correlations encoded into $F_{\psi}(\lambda_{r},\lambda_{l})$ one needs, first of all, a frequency-resolved detection scheme. Second, we suggest to measure the FCS using the philosophy employed in \cite{Hoffer} in the cold atom context. Instead of studying the moments of distribution to a certain order one can directly plot histograms of measurement outcomes using homodyne detection scheme. Recent experiments on tomography of quantum states in multi-mode transmission lines coupled to q-bits have employed similar method \cite{Eichler} in the microwave domain. Typical interaction strength can be varied between 50 MHz -300MHz while the typical transmission line frequencies are of the order of 500MHz and environmental losses are typically small.  We note that changing the pulse's duration one can effectively change the coupling strength $\gamma$. The line width of the source employed in \cite{Eichler} for microwave transmission lines corresponds to the very large pulse duration limit (large $L$) where the universal effects described above can be directly observed.  A selective enhancement of the $n$-particle scattering effects can be achieved via balancing between parameters of the initial state and field-emitter coupling strength.

{\it Conclusion}. In conclusion, we have exactly computed for the first time  a full counting statistics of a fundamental problem in quantum optics - a pulse of coherent light propagating in a multi-mode channel and interacting with a two-level system. These results allow for the quantitative determination of the effects of many-body correlations coming from the interplay of photons' dispersion, interaction of photons with emitter and physical parameters of a pulse (duration and average number of photons). By tuning the parameters of a pulse one can selectively access $n$-particle correlations. For generic initial state the limiting form of statistics is given by a bimodal distribution. Effects described here can be directly observed experimentally with existing setups. 

{\it Acknowledgements}. We benefited a lot from discussions with A. Fedorov, A. Komnik, M. Laakso and G. Morigi. M. R. and V. G. are supported by Swiss NSF. M. P. acknowledges the financial support from DFG-FG 723. V.G thanks KITP for hospitality.

\newpage
\clearpage
\onecolumngrid

\renewcommand{\theequation}{S\arabic{equation}} 
\renewcommand{\thepage}{S\arabic{page}} 
\renewcommand{\thesection}{S\arabic{section}}  
\renewcommand{\thetable}{S\arabic{table}}  
\renewcommand{\thefigure}{S\arabic{figure}}
\renewcommand{\bibnumfmt}[1]{[{\normalfont S#1}]}
\setcounter{page}{0}
\setcounter{equation}{0}

\begin{center}
{\bf \Large Supplementary Material}
\end{center}

Here we outline essential technical details of our derivations and present some additional results. 
\section{Theoretical model and scattering formalism}

\subsection{Theoretical model}
Our model is motivated by current experiments in nanophotonics where a one-dimensional (1D) transmission line is coupled to an emitter (an artificial atom, a quantum dot, a NV center) or cold atoms are coupled to evanescent modes. More generally, the model in question can be considered as a basic model of quantum optics, since quasi-1D models effectively emerge when the scattering is restricted to the s-wave channel. 

Its derivation relies on a number of approximations which are customary in quantum optics: (i) a dipole approximation, (ii) the rotating wave approximation (RWA), and (iii) Born-Markov approximation. In addition, we linearize the photonic spectrum around some appropriately chosen working frequency $\Omega_{0}$ which is commensurate with the emitter's transition frequency $\Omega$. Under these assumptions [except for (ii)] we obtain an effective low-energy Hamiltonian
\beq
H=\sum_{\xi=r,l}\int dk(\Omega_{0}+\xi k)a^{\dag}_{\xi k}a_{\xi k}+\frac{\Omega}{2}\sigma^{z}+g_0 \sum_{\xi=r,l}\int dk (a_{\xi k}^{\dag}+a_{\xi k})(\sigma^{+}+\sigma^{-}),
\eeq
featuring the two-branch linear dispersion with right- ($\xi=r=+$) and left- ($\xi=l=-$) propagating modes. To implement the RWA in a systematic way, we first perform the gauge transformation $H \to U^{\dag} H U  + i (d U^{\dag}/ dt) U$ with 
\beq
U=\exp\left[-i\Omega_{0}t\left(\sum_{\xi=r,l}\int dk \, a^{\dag}_{\xi k} a_{\xi k}+\frac{\sigma^z}{2} \right)\right],
\eeq
which leads us the Hamiltonian
\beq
H&=&\sum_{\xi=r,l}\int dk \, \xi \, k \, a^{\dag}_{\xi k}a_{\xi k}+\frac{\Delta}{2}\sigma^{z}+g_0 \sum_{\xi=r,l}\int dk (a_{\xi k}^{\dag}\sigma^{-}+a_{\xi k}\sigma^{+}) \label{start-Ham} \\
&+& g_0 \sum_{\xi=r,l}\int dk (a_{\xi k}^{\dag}\sigma^{+} e^{2i\Omega_{0}t}+a_{\xi k}\sigma^{-} e^{-2i\Omega_{0}t}),
\label{osc-terms}
\eeq
where $\Delta=\Omega-\Omega_{0}$. As soon as $g_0^2/\Omega_0 \ll 1$, the oscillating terms \eqref{osc-terms} can be treated as a perturbation. In zeroth order they are simply neglected, what is equivalent to the RWA.

\subsection{Definition of wave packet field operators}

In order to define a coherent state we need field operators annihilating/creating states normalized to unity. The field operators $a_{k}$ and $a^{\dag}_{k}$, which are the Fourier transforms of $a (x)$ and $a^{\dag} (x)$, do not suit for this purpose, as they fulfill the commutation relation $[a_k, a^{\dagger}_{k'}] = \delta (k-k')$. To circumvent this difficulty, we construct wave packet field operators
\be
b_k = \frac{1}{\sqrt{L}} \int_{-L/2}^{L/2} d x a (x ) e^{- i k x}, \quad b_k^{\dagger} = \frac{1}{\sqrt{L}} \int_{-L/2}^{L/2} d x a^{\dagger} (x ) e^{i k x},
\label{op_c}
\ee
which do satisfy the desired commutation relation $[b_{k_0}, b_{k_0}^{\dagger}] =1$.
They annihilate/create photons spatially localized on a finite interval of the length $L$. With their help we are able to introduce the coherent state
\be
| \alpha_{k_0} \rangle  = e^{\alpha_{k_0} b_{k_0}^{\dagger} - \alpha_{k_0}^* b_{k_0}} | 0 \rangle = e^{-|\alpha_{k_0} |^2/2} \sum_{n=0}^{\infty} \frac{\alpha_{k_0}^{n}  (b_{k_0}^{\dagger})^n}{n!} |0 \rangle , 
\label{init_coherent}
\ee
possessing the proper normalization $\langle \alpha_{k_0} | \alpha_{k_0} \rangle =1$.

\subsection{Transformation to the ``even-odd'' basis}

Due to energy independence of the coupling constant $g_0$ (which is the part of  Born-Markov approximation), one can decouple the model defined by \eqref{start-Ham} into two sectors. To this end, one introduces \emph{even} (or symmetric) and \emph{odd} (or antisymmetric) combinations of fields corresponding to the same energy
\be
a_{ek} = \frac{a_{rk} + a_{l, -k}}{\sqrt{2}}, \quad a_{ok} = \frac{a_{rk}-  a_{l, -k}}{\sqrt{2}} .
\ee
By virtue of this canonical transformation the Hamiltonian \eqref{start-Ham} turns into a sum of the two terms, $H=H_e + H_o$, defined by
\beq
H_e &=& \int dk \left[ k \, a^\dag_{e k} a_{e k} 
+ g  \left( a^\dag_{e k}\sigma^-  + a_{e k}\sigma^+ \right) \right]
+ \frac{\Delta}{2} \sigma^z , \\
H_o &=&  \int dk \, k \,  a^\dag_{ok} a_{ok},
\eeq
where $g=g_0 \sqrt{2}$. Note that the odd Hamiltonian $H_{o}$ is noninteracting, and therefore  odd modes do not scatter off a local emitter ($S_o \equiv \mathbbm{1}$). The even Hamiltonian $H_{e}$ can be interpreted in terms of a chiral model with a single branch of the linear dispersion. Scattering in chiral models has been studied Ref.~\cite{PGnjp} for arbitrary initial states including the coherent state. If  an initial state of the model \eqref{start-Ham} defined in the original right-left basis admits a decomposition into a product state in the even-odd basis, and this is indeed the case for the coherent state
\be
e^{\alpha b^{\dag}_{r, k_0} -|\alpha|^2/2} | 0 \rangle = 
e^{(\alpha/\sqrt{2}) b^{\dag}_{e,k_0} -|\alpha|^2/4} e^{(\alpha/\sqrt{2}) b^{\dag}_{o,k_0} - |\alpha|^2/4} | 0 \rangle \equiv  |  \alpha/\sqrt{2} \rangle_e \otimes | \alpha/\sqrt{2} \rangle_o ,
\label{init-state}
\ee
then it suffices to apply the results of \cite{PGnjp} in order to find the scattering state $S_e | \alpha / \sqrt{2} \rangle_e$. All expressions necessary for this purpose are presented below.

\subsection{Scattering of the coherent state in the chiral model}

For the chiral model the outgoing state resulting from scattering of the initially coherent state $| \alpha_{k_0} \rangle_{\mathrm{ch}}$ (in the mode $k_0$) has been constructed in Eq.~(134) of Ref.~\cite{PGnjp}. For completeness, we quote it here as well
\beq
& & | \mathrm{out}, \alpha_{k_0} \rangle_{\mathrm{ch}} = S_{\mathrm{ch}} | \alpha_{k_0} \rangle_{\mathrm{ch}} , \label{fin_coherent} \\
& &  S_{\mathrm{ch}} = 1 + \sum_{n=1}^{\infty} \lambda^n \int d x_1 \ldots d x_n e^{i k_0 (x_1 + \ldots +x_n)}  a^{\dagger} (x_1 ) \ldots a^{\dagger} (x_n) \nonumber \\
& & \times \left[ \Theta
(L/2 > x_n > \ldots > x_2 >x_1 > - L/2) \prod_{j=1}^n \left( 1 - e^{i (k_0-\Delta_g) \delta x_{j+1}}\right) \right. \nonumber \\
&+& \left. \Theta
(L/2 > x_n > \ldots > x_2 > - L/2 >x_1)  e^{- i (k_0-\Delta_g) (x_1 +L/2)} \left( 1- e^{i (k_0-\Delta_g) (x_2 +L/2)} \right) \prod_{j=2}^n \left( 1 - e^{i (k_0-\Delta_g) \delta x_{j+1}}\right)\right], \nonumber\\
\label{Smatr}
\eeq
where $\delta x_{j}=x_{j}-x_{j-1}>0$, $x_{n+1} \equiv L/2$, as well as $\Delta_g = \Delta - i \pi g^2$ and $\lambda = - 2 \pi i g^2 \alpha_{k_0}/[(k_0 - \Delta_g) \sqrt{L}]$.

Let us now evaluate the factor 
\be
\mathcal{F} (v_{k_0}^*) = 
\langle 0 |   e^{v_{k_0}^* b_{k_0}}  | \mathrm{out}, \alpha_{k_0} \rangle_{\mathrm{ch}} =  e^{-|\alpha_{k_0}|^2/2 + v_{k_0}^* \alpha_{k_0}} \langle 0 | e^{v_{k_0}^* b_{k_0}} S_{\mathrm{ch}}  e^{- v_{k_0}^* b_{k_0}}   | 0 \rangle 
\label{Sfactor1}
\ee
for arbitrary $v_{k_0}^*$.  Permuting $e^{v_{k_0}^* b_{k_0}}$ in \eqref{Sfactor1} to the right produces the shift of fields $a^{\dagger} (x_i) \to a^{\dagger} (x_i) + \frac{v_{k_0}^*}{\sqrt{L}} e^{-i k_0 x_i}$ spanning the scattering matrix $S_{\mathrm{ch}}$. Reorganizing the resulting series for $S[v_{k_0}^*] =  e^{v_{k_0}^* b_{k_0}} S_{\mathrm{ch}} e^{-v_{k_0}^* b_{k_0}}$, we obtain the following expression 
\beq
S[v_{k_0}^*] = S^a [v_{k_0}^*; L/2, -L/2] + S^b [v_{k_0}^*; L/2, -L/2] \frac{\lambda}{g \sqrt{2 \pi}} A_0^{\dagger},
\label{S-factor}
\eeq
where
\be
A_0^{\dagger} = g \sqrt{2 \pi} \int_{-\infty}^{-L/2} d x_0 e^{i k_0 x_0} e^{- i (k_0-\Delta_g) (L/2 +x_0)} a^{\dagger} (x_0), \quad \langle 0 | A_0 A_0^{\dagger}  | 0 \rangle =1 ;
\ee
and
\beq
S^a [v_{k_0}^* ; x,y]  &=& \tilde{d}_v (x-y) +\sum_{n=1}^{\infty} \lambda^n \int d x_n \ldots d x_1 \Theta (x > x_n > \ldots > x_1 > y \geq -L/2) \nonumber \\
& \times & d_v (x - x_n) a^{\dagger} (x_n) e^{i k_0 x_n} d_v (x_n - x_{n-1}) a^{\dagger} (x_{n-1}) e^{i k_0 x_{n-1}} \ldots a^{\dagger} (x_1 ) e^{i k_0 x_1} \tilde{d}_v (x_1 -y),  \label{Sa} \\
S^b [v_{k_0}^* ; x,y] &=& d_v (x-y) +\sum_{n=1}^{\infty} \lambda^n \int d x_n \ldots d x_1 \Theta (x > x_n > \ldots > x_1 > y \geq -L/2) \nonumber \\
& \times & d_v (x - x_n) a^{\dagger} (x_n) e^{i k_0 x_n} d_v (x_n - x_{n-1}) a^{\dagger} (x_{n-1}) e^{i k_0 x_{n-1}} \ldots a^{\dagger} (x_1 ) e^{i k_0 x_1} d_v (x_1 -y). \label{Sb}
\eeq
The operators \eqref{Sa} and \eqref{Sb} are defined with help of the kernels
\beq
d_v (x) &=& - \frac{p_+ + p_-}{p_+ - p_-} \left[ e^{- i p_+ x} - e^{- i p_- x}\right] , \\
\tilde{d}_v (x) &=&  - \frac{p_-}{p_+ - p_-}  e^{- i p_+ x} + \frac{p_+ }{p_+ - p_-} e^{- i p_- x},
\eeq
where $p_{\pm} = p_{\pm} (v_{k_0}^*)$ are the roots of the quadratic equation
\be
p^2 + (k_0 - \Delta_g) p -2 \pi g^2 \alpha_{k_0} v_{k_0}^*/L =0.
\ee

From this representation we straightforwardly get $\mathcal{F} (v_{k_0}^*) = \tilde{d}_v (L) \equiv \tilde{d}_v (L, \alpha_{k_0})$.

\section{Computing FCS}

Due to factorization of the initial state \eqref{init-state} and of the scattering matrix $S=S_e S_o = S_e \mathbbm{1}$ in the even-odd basis, we also obtain the scattering state in the product form 
\be
| \mathrm{out} \rangle =S | \mathrm{in} \rangle = | \mathrm{out}, \alpha/\sqrt{2} \rangle_e \otimes | \alpha /\sqrt{2} \rangle_o
\label{out-state} 
\ee

As we want to measure moments of the operators $N_{\xi}$ ($\xi=r,l$), we start our consideration of the FCS from
\beq
e^{i \lambda_{\xi} N_{\xi} } &=&
1+ \sum_{n=0}^{\infty} \frac{1}{n !} \left( e^{i \lambda_{\xi} n} -1\right) b_{\xi}^{\dagger n}  | 0 \rangle \langle 0 | b_{\xi}^n  \nonumber \\
&=& 1+ \sum_{n=0}^{\infty} \frac{1}{n !} \left( e^{i \lambda_{\xi} n} -1\right) \left( \frac{b_{e,k_0}^{\dagger } + \xi b_{o,k_0}^{\dagger}}{\sqrt{2}} \right)^n | 0 \rangle \langle 0 | \left( \frac{b_{e, k_0} + \xi b_{o,k_0}}{\sqrt{2}} \right)^n .
\eeq
It follows that
\beq
e^{i \lambda_{r} N_{r} +i \lambda_l N_l} &=&
1+ \sum_{n=0}^{\infty} \frac{1}{n !} \left( e^{i \lambda_{r} n} -1\right) b_{r}^{\dagger n}  | 0 \rangle \langle 0 | b_{r}^n + \sum_{m=0}^{\infty} \frac{1}{m !} \left( e^{i \lambda_{l} m} -1\right) b_{l}^{\dagger m}  | 0 \rangle \langle 0 | b_{l}^m  \nonumber \\
&+& \sum_{n,m=0}^{\infty} \frac{1}{n! m!} \left( e^{i \lambda_{r} n} -1\right)\left( e^{i \lambda_{l} m} -1\right) b_{r}^{\dagger n} b_{l}^{\dagger m}  | 0 \rangle \langle 0 | b_{l}^m b_{r}^n
\label{exp-iden} 
\eeq

Representing $b_{\xi} = (b_{e,k_0} + \xi b_{o,k_0})/\sqrt{2}$ and calculating the quantum average of \eqref{exp-iden} in the outgoing state \eqref{out-state}, we obatin
\beq
F (\lambda_r , \lambda_l) &=& 1+ \sum_{n=0}^{\infty} \frac{1}{n!}  \left( e^{i \lambda_{r} n} -1\right) |S_{n0}|^2 + \sum_{m=0}^{\infty} \frac{1}{m!} \left( e^{i \lambda_{l} m} -1\right) |S_{0m}|^2 \nonumber \\
&+& \sum_{n,m=0}^{\infty}  \frac{1}{n! m!} \left( e^{i \lambda_{r} n} -1\right) \left( e^{i \lambda_{l} m} -1\right) |S_{nm}|^2,
\eeq
where
\beq
S_{nm} &=&   e^{-|\alpha |^2/4} \langle 0 | \left( \frac{b_{e,k_0} + \alpha/\sqrt{2} }{\sqrt{2}} \right)^n \left( \frac{b_{e,k_0} - \alpha/\sqrt{2} }{\sqrt{2}} \right)^m |
\mathrm{out} , \alpha /\sqrt{2} \rangle_e \nonumber \\
&=&  e^{-|\alpha |^2/4} \frac{\partial^n}{\partial (v_r^* )^n} \frac{\partial^m}{\partial (v_l^* )^m} \left\{ \langle 0 | \exp \left[ v_r^* \left( \frac{b_{e,k_0} +  \alpha/\sqrt{2} }{\sqrt{2}} \right) + v_l^* \left( \frac{b_{e,k_0} - \alpha/\sqrt{2} }{\sqrt{2}} \right)\right] |
\mathrm{out} , \alpha /\sqrt{2} \rangle_e \right\}_{v_{\xi}^*=0} \nonumber \\
&=&   e^{-|\alpha |^2/4} \frac{\partial^n}{\partial (v_r^* )^n} \frac{\partial^m}{\partial (v_l^* )^m} \left\{ e^{ (v_r^* - v_l^*) \alpha /2} \langle 0| e^{(v^*_r + v^*_l)b_{e,k_0}/\sqrt{2}} | \mathrm{out} , \alpha /\sqrt{2} \rangle_e \right\}_{v_{\xi}^*=0} \nonumber \\
&=&  e^{-|\alpha |^2/2} \frac{\partial^n}{\partial (v_r^* )^n} \frac{\partial^m}{\partial (v_l^* )^m}\left\{ e^{v_r^* \alpha } \tilde{d}_{(v_r + v_l)/\sqrt{2}} (L, \alpha /\sqrt{2}) \right\}_{v_{\xi}^*=0} .
\eeq
Introducing new variables $w_{\pm} = \frac{\alpha}{2} (v^*_r \pm v^*_l) $, we obtain
\beq
S_{nm} = e^{-|\alpha |^2/2}  \left(\frac{\alpha}{2} \right)^{n+m} \left( \frac{\partial}{\partial w_+} + \frac{\partial}{\partial w_-}\right)^n  \left( \frac{\partial}{\partial w_+} - \frac{\partial}{\partial w_-}\right)^m  \left\{ e^{w_+ + w_- } \tilde{d}_{w_+} (L) \right\}_{w_{\pm} =0} 
\equiv  e^{-|\alpha |^2/2} \alpha^{n+m} s_{nm}  ,
\eeq
where
\be
s_{nm} = \frac{1}{2^{n+m}} \left( \frac{\partial}{\partial w} + 2\right)^n  \left( \frac{\partial}{\partial w} \right)^m   \tilde{d}_{w} (L) \bigg|_{w =0} ,
\quad w \equiv w_+ ,
\label{s_n-coeff}
\ee
and
\beq
\tilde{d}_{w} (L) = e^{i (\delta + i \gamma)/2} \left[\cos (\kappa L/2) -i  (\delta + i \gamma ) \frac{\sin (\kappa L/2)}{\kappa L}  \right], \quad
\kappa L  = \sqrt{(\delta +i\gamma)^2 + 8 \gamma w } .
\label{d-w}
\eeq
Then
\beq
F_{\alpha} (\lambda_r , \lambda_l)
&=& 1 +\sum_{n=0}^{\infty} (e^{i \lambda_r n}-1) p_{\alpha} (n) |s_{n0} |^2 + \sum_{m=0}^{\infty} (e^{i \lambda_l m}-1) p_{\alpha} (m) |s_{0m} |^2 \nonumber \\
&+& e^{|\alpha|^2} \sum_{n,m=0}^{\infty} (e^{i \lambda_r n}-1)(e^{i \lambda_l m}-1) p_{\alpha} (n) p_{\alpha} (m)  |s_{nm} |^2 .
\eeq

\begin{figure}
\begin{center}
    \includegraphics[width=0.48\textwidth]{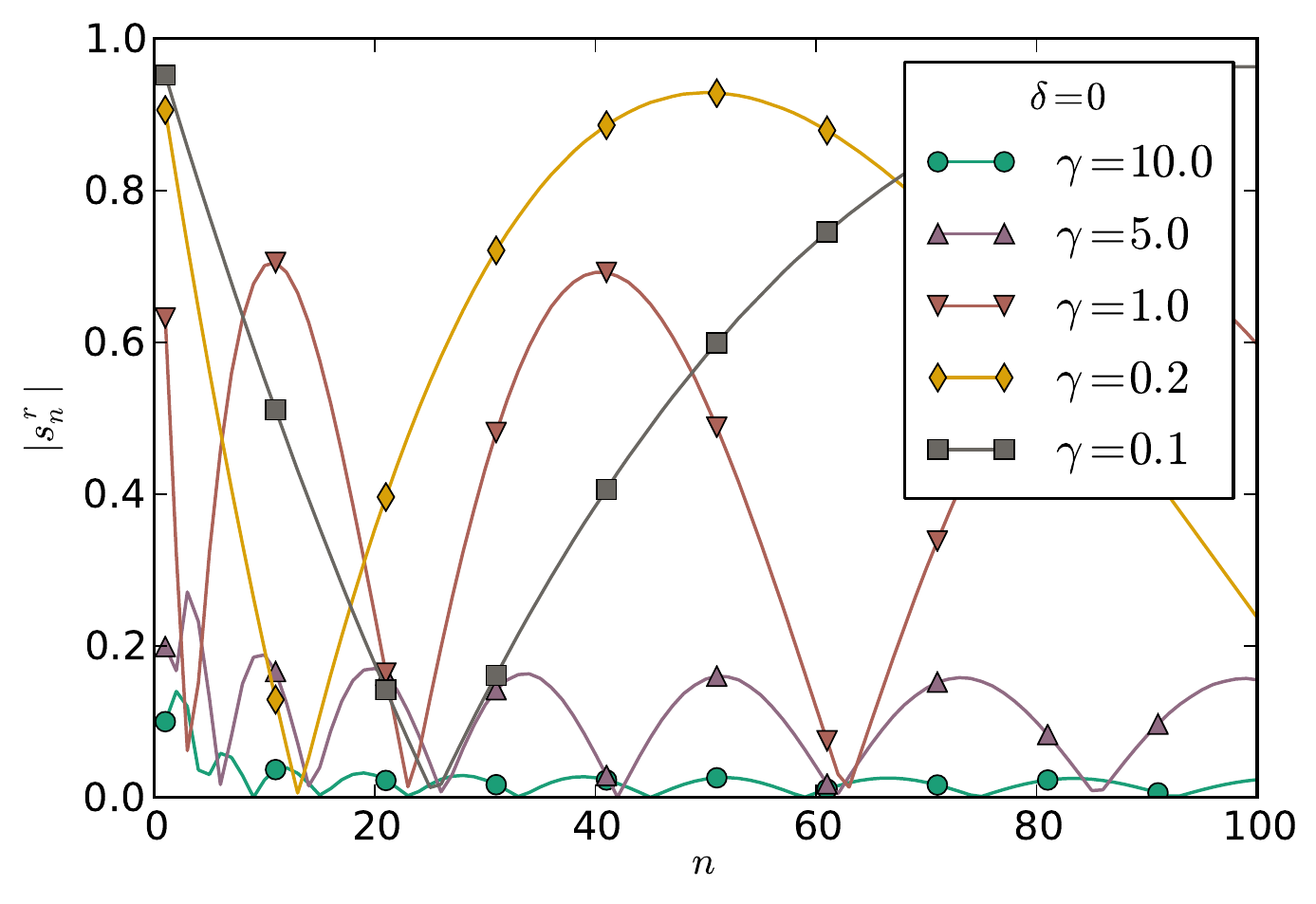}
    \includegraphics[width=0.48\textwidth]{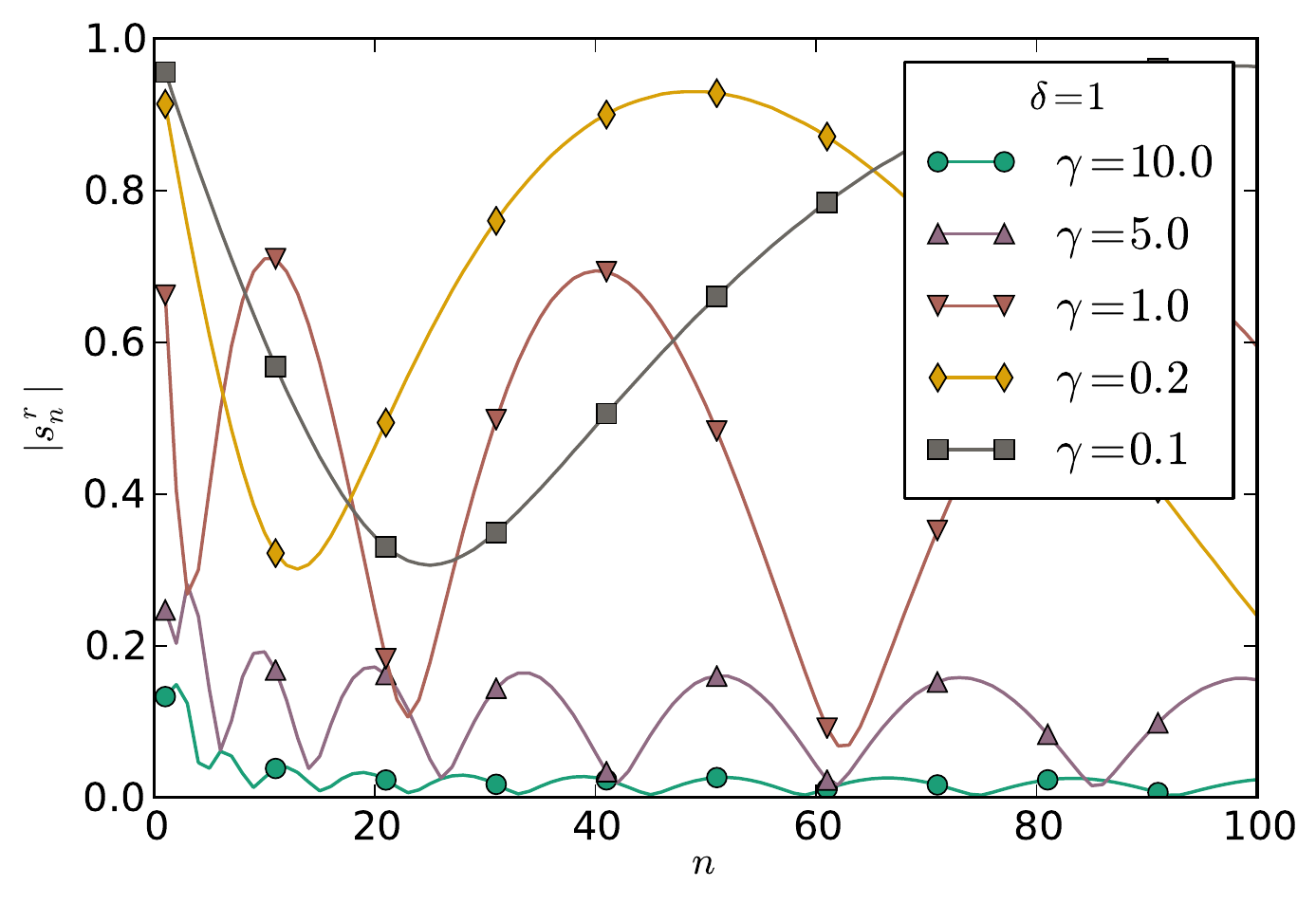}
    \includegraphics[width=0.48\textwidth]{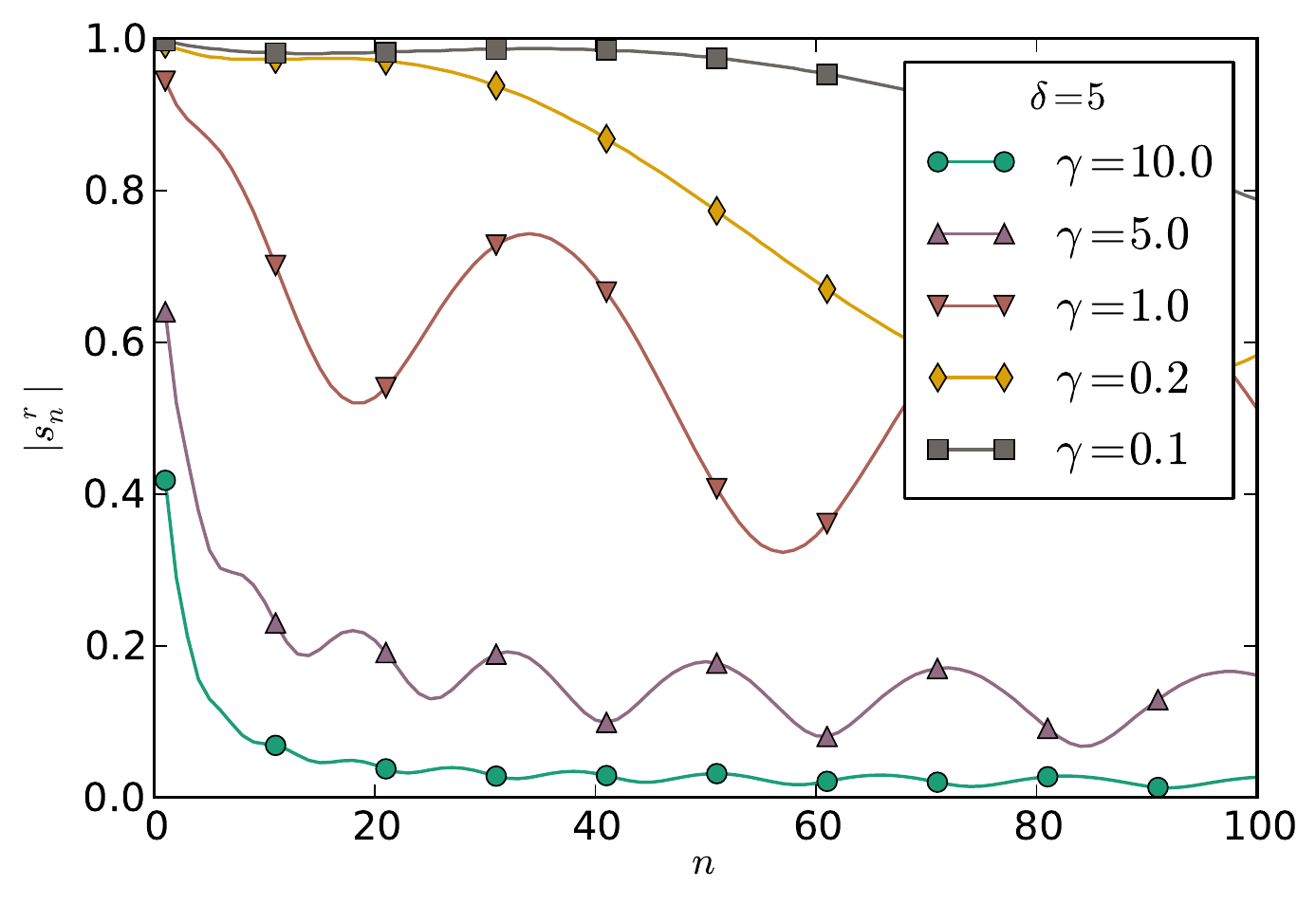}
\end{center}
\caption{The dependence of magnitudes of coefficients $s_n^{r,l}$, introduced at
Eq.~\eqref{eq:DefinitionOfFalpha} on $n$ at various coupling
constants~$\gamma$ and  detunings~$\delta$.  }
\label{fig:sRvarGamma}
\end{figure}

\begin{figure}
\begin{center}
    \includegraphics[width=0.48\textwidth]{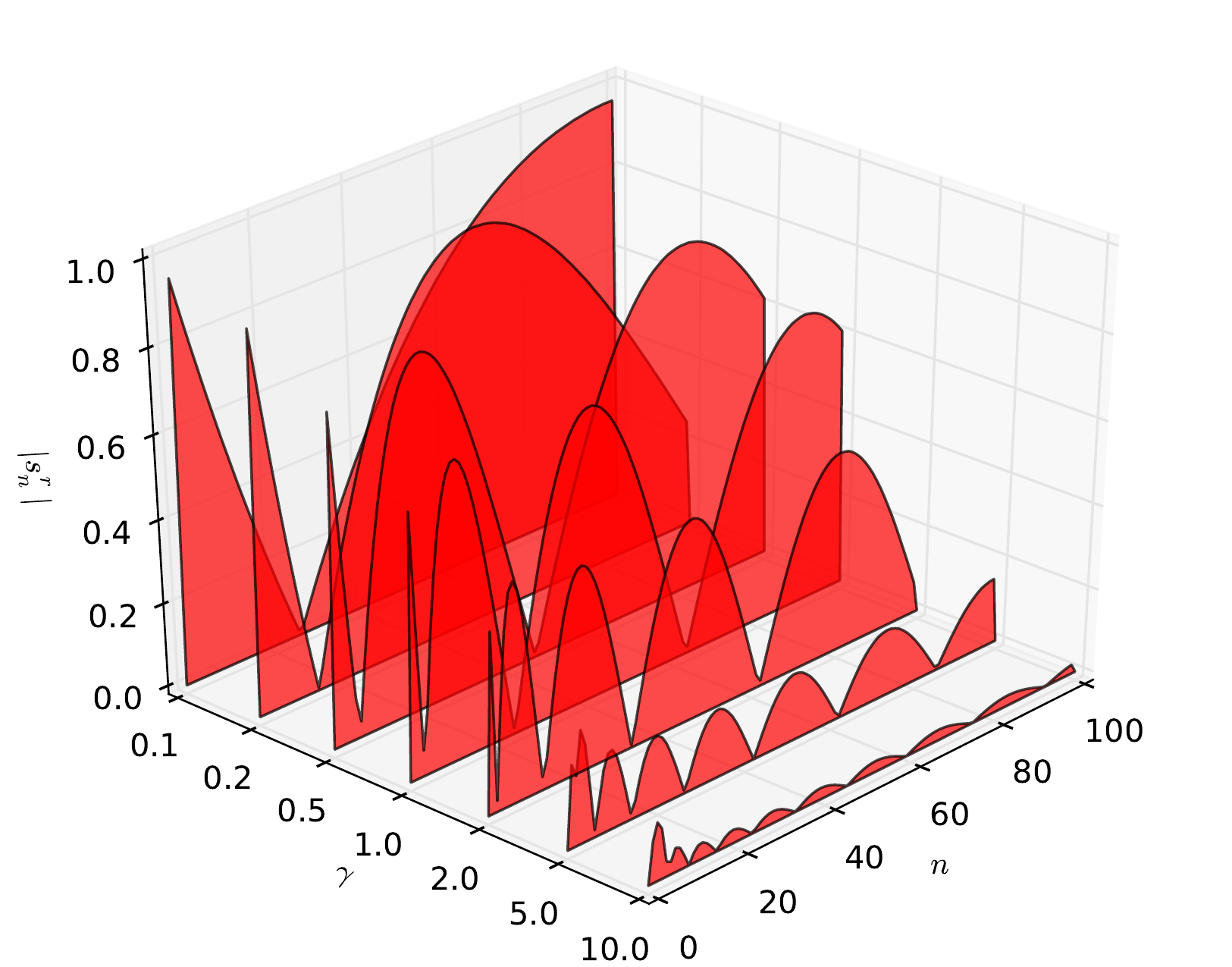}
\end{center}
\caption{The dependence of magnitudes of coefficients $s_n^{r,l}$, introduced at
Eq.~\eqref{eq:DefinitionOfFalpha} on $n$ at various coupling
constants~$\gamma$, while the detuning~$\delta=0$. 
This is another representation of same data from Fig.~\ref{fig:sRvarGamma}.
}
\label{fig:sRvar3D}
\end{figure}

Introducing  notations $\rho= (\delta+i\gamma)/2$ and $t=-\gamma w/\rho$, and using an expression for the generating function of  the spherical Bessel functions 
\beq
\frac{1}{z}\cos(\sqrt{\rho^{2}-2\rho t}) = \sum_{n=0}^{\infty}\frac{t^{n}}{n!}j_{n-1}(\rho),
\eeq 
we cast \eqref{d-w} to
\beq
\tilde{d}_{w}(L) = \sum_{n=0}^{\infty} \frac{t^{n}}{n!}c_{n}(\rho) ,
\quad
c_{n}(\rho)=\rho e^{i \rho} [ j_{n-1}(\rho)-ij_{n}(\rho)] .
\eeq
and from Eq.~\eqref{s_n-coeff} explicitly calculate
\beq
s_{nm} &=& \sum_{p=0}^n \binom{n}{p} \left(\frac{-\gamma}{2\rho}\right)^{p+m} c_{p+m}(\rho).
\eeq
Note that these coefficients do not depend on $\alpha$.
For illustration we plot the coefficients $s_n^r$ and $s_n^l$ in Figs.~\ref{fig:sRvarGamma} and
\ref{fig:sRvar3D}.

At large $|\rho| \gg 1$ we use an asymptotic expression for $c_n \approx i^n$ and observe a
factorization of the coefficients $s_{nm} = t^n r^m$, where $t=\delta/(\delta +i \gamma)$ and $r =
-i \gamma/(\delta + i \gamma)$ are transmission and reflection amplitudes of the single photon scattering, respectively.

Then, in the limit $L \to \infty$ (continuous laser radiation) we obtain
\beq
F_{\alpha} (\lambda_r , \lambda_l)
&=& 1 +\sum_{n=0}^{\infty} (e^{i \lambda_r n}-1) p_{\alpha} (n) T^n + \sum_{m=0}^{\infty} (e^{i \lambda_l m}-1) p_{\alpha} (m) R^m \nonumber \\
&+& e^{\bar{N}} \sum_{n,m=0}^{\infty} (e^{i \lambda_r n}-1)(e^{i \lambda_l m}-1) p_{\alpha} (n) p_{\alpha} (m)  T^n R^m .
\label{cont-limit}
\eeq

After a simple transformation we arrive at the explicit formula
\beq
F_{\alpha} (\lambda_r , \lambda_l)
&=& 1 +e^{-\bar{N}} \sum_{n=0}^{\infty} \frac{(\bar{N} T)^n}{n!} (e^{i \lambda_r n}-1) + e^{-\bar{N} } \sum_{m=0}^{\infty} \frac{(\bar{N} R )^m}{m!} (e^{i \lambda_l m}-1) \nonumber \\
&+& e^{-\bar{N}} \sum_{n,m=0}^{\infty} \frac{(\bar{N} T)^n (\bar{N} R)^m}{n! m!} (e^{i \lambda_r n}-1)(e^{i \lambda_l m}-1) , \nonumber \\
&=& 1 + e^{-\bar{N}} \left( e^{\bar{N} T e^{i \lambda_r}} -e^{\bar{N} T}\right)
+ e^{-\bar{N}} \left( e^{\bar{N} R e^{i \lambda_l}} -e^{\bar{N} R}\right)+ e^{-\bar{N}} \left( e^{\bar{N} T e^{i \lambda_r}} -e^{\bar{N} T}\right)
 \left( e^{\bar{N} R e^{i \lambda_l}} -e^{\bar{N} R}\right)
\eeq
where $T= |t|^2$ and $R=|r|^2 = 1-T$.

Considering the case $\lambda_l=0$ we obtain the bimodal distribution for transmitted  photons
\beq
F (\lambda_r , 0) = 1-e^{-R \bar{N}} + e^{- R \bar{N}} e^{T \bar{N} (e^{i \lambda_r}-1)},
\eeq 
which interpolates between Poissonian and zero occupancy distributions. Analogously for reflected photons ($\lambda_r = 0$).

In the continuous laser radiation limit we can generalize \eqref{cont-limit} for arbitrary
initial state $|\psi_{k_0} \rangle = \sum_{n=0}^{\infty} \psi_n |n \rangle_{k_0}$ created in the mode $k_0$ with the distribution of photon numbers $p (n) = |\psi_n|^2$. This becomes possible due the existence of the identity resolution 
\be
1 = \int \frac{d^2 \alpha}{\pi} | \alpha \rangle \langle \alpha | .
\ee
Then,
\be
S_{nm}  = s_{nm} \int \frac{d^2 \alpha}{\pi}  e^{-|\alpha|^2/2} \alpha^{n+m} \langle \alpha | \psi_{k_0} \rangle = s_{nm} \int \frac{d^2 \alpha}{\pi}  e^{-|\alpha|^2} \alpha^{n+m} \sum_{l=0}^{\infty} \frac{\alpha^{*l}}{\sqrt{l!}} \psi_l =   \sqrt{(n+m)!} \psi_{n+m} s_{nm}.
\ee
and
\beq
F_{\psi} (\lambda_r , \lambda_l) &=& 1 + \sum_{n=0}^{\infty} \left( e^{i \lambda_r n} -1\right) p (n) T^n  + \sum_{m=0}^{\infty} \left( e^{i \lambda_l m} -1\right) p (m) R^m \nonumber \\
&+& \sum_{n,m=0}^{\infty}  \binom{n+m}{n} \left( e^{i \lambda_{r} n} -1\right) \left( e^{i \lambda_{l} m} -1\right) p (n+m) T^n R^m,
\eeq
Thus, for the Fock state with $N$ photons $p (n) = \delta_{n,N}$ we obtain
\be
F _{N}(\lambda_r , 0) = 1+ \left( e^{i \lambda_r N} -1\right) T^N,
\ee
which means that either all $N$ photons are coherently transmitted with the probability $T^N$ or all of them are reflected with the probability $1 -T^N$.

For the squeezed state $|\zeta\rangle=S(\zeta)|0\rangle$ where $S(\zeta)=\exp(\frac{1}{2}\zeta^{*}b^{2}_{R}-\frac{1}{2}\zeta (b^{\dag}_{R})^{2})$ and $\zeta=|\zeta|e^{i\theta}$ the coefficients  $\psi_{2n}=(2n-1)!!(-e^{i\theta}\tanh|\zeta|)^{n}/\sqrt{(2n)!\cosh|\zeta|}$, while $\psi_{2n+1}=0$. Then 
\beq
F_{\zeta}(\lambda_r ,0)=(1-d_{\zeta})\delta_{n,0}+d_{\zeta} p_{\zeta'}(n)
\eeq
where $d_{\zeta}=\cosh|\zeta'|/\cosh|\zeta|$, and $p_{\zeta'}(n)$ is obtained from $p_{\zeta}(n)$ by replacing $|\zeta|$ by $|\zeta|'=\mbox{arctanh}(T^2\tanh|\zeta|)$.

\section{Moments of the distribution}

Here we present some additional results on the mean number of photons remaining in the mode $k_0$ in the transmission and reflection channels as well as probability distribution function for different set of parameters. 
\begin{figure}[h]
\includegraphics[width=0.48\textwidth]{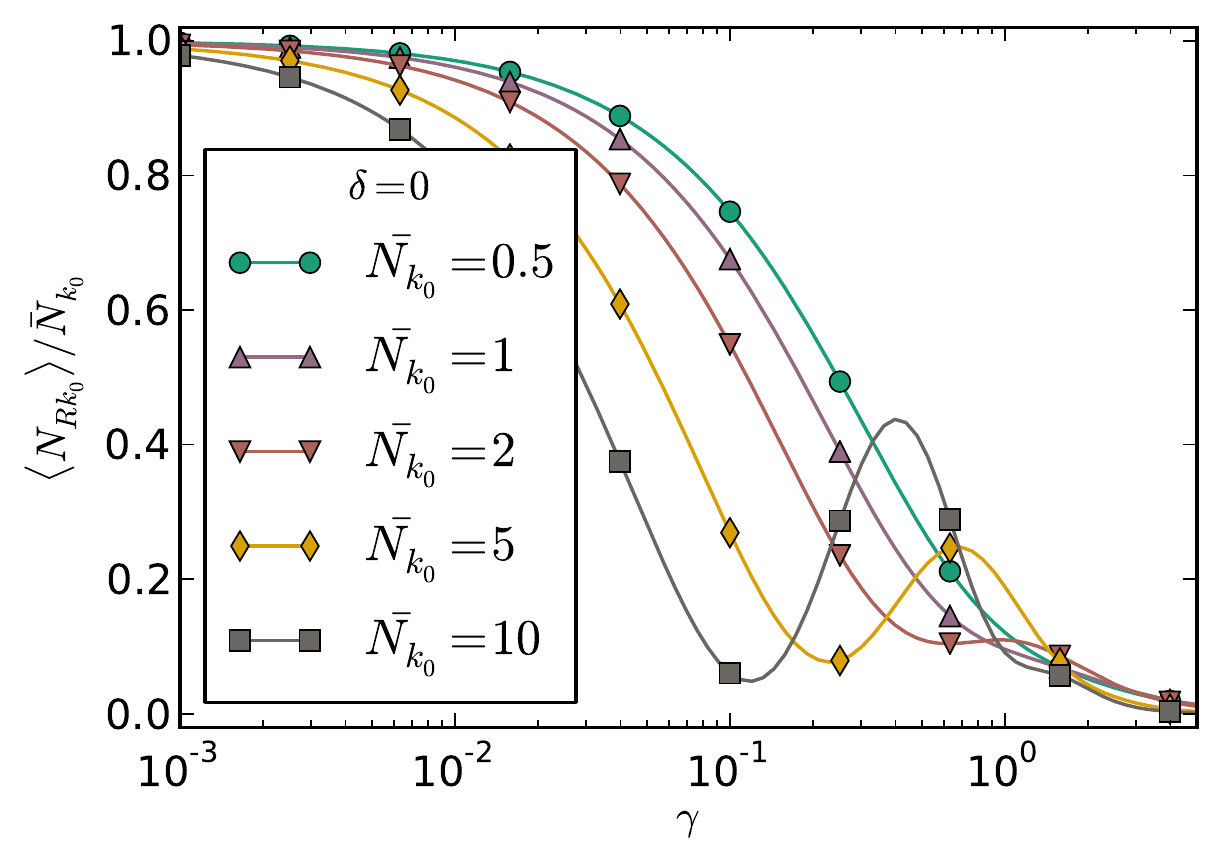}
\includegraphics[width=0.48\textwidth]{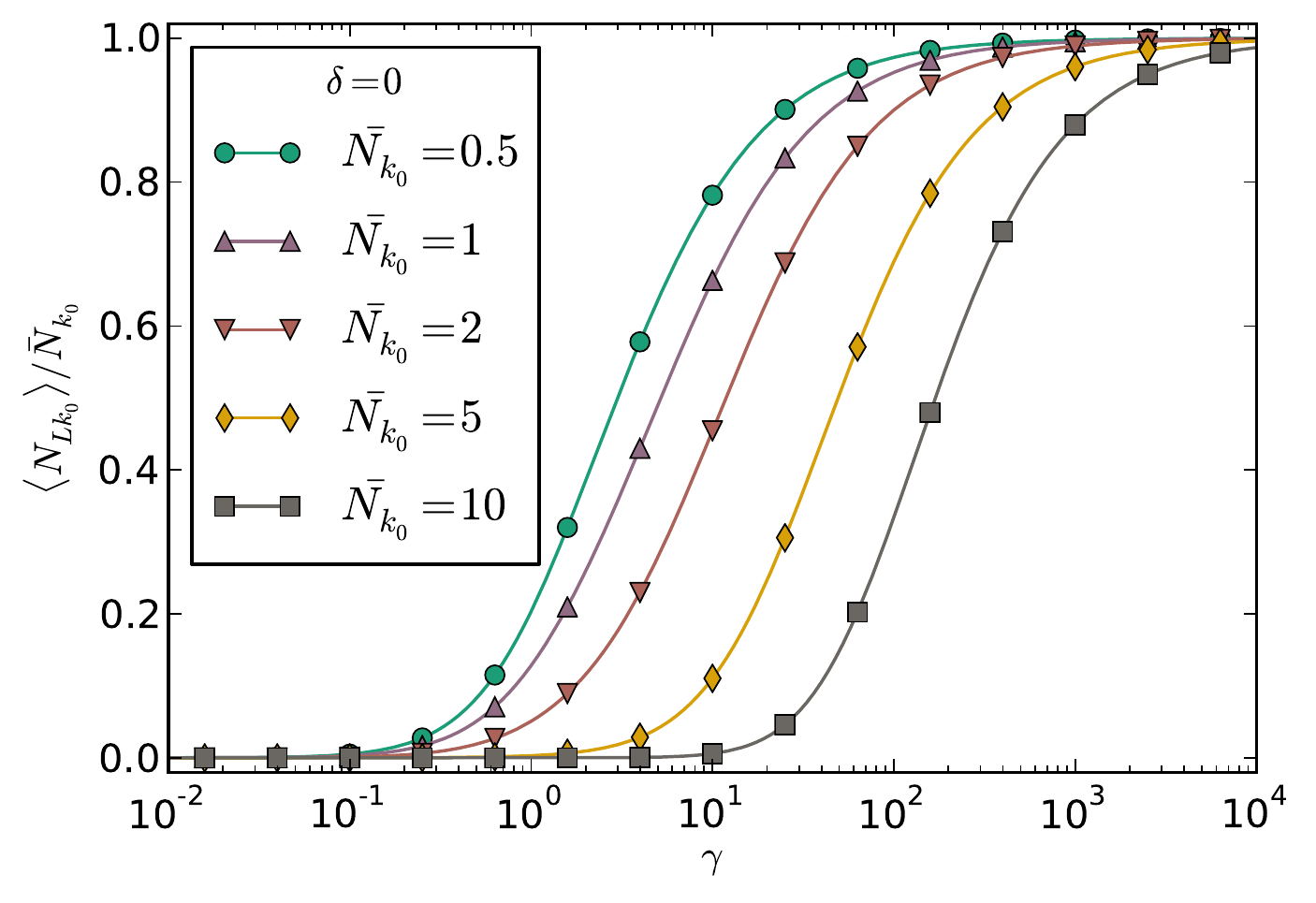}
\includegraphics[width=0.48\textwidth]{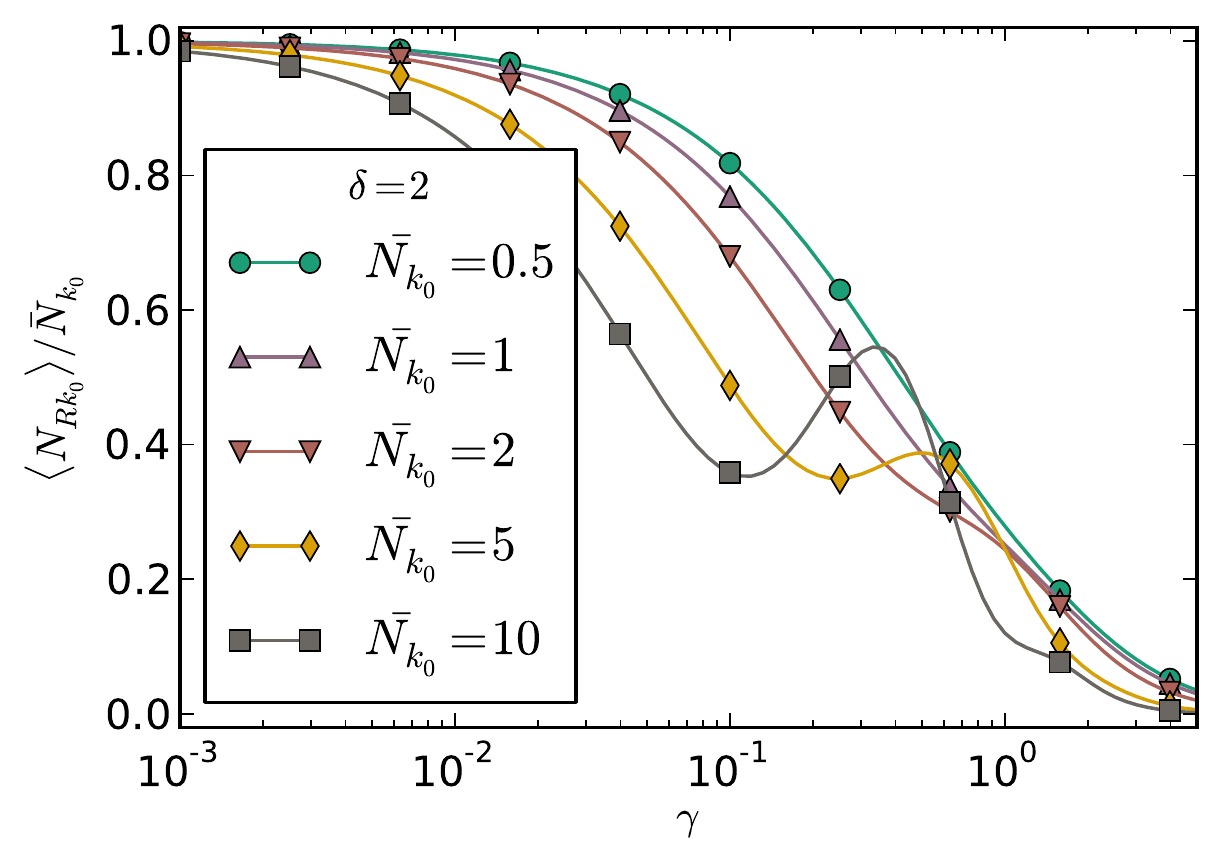}
\includegraphics[width=0.48\textwidth]{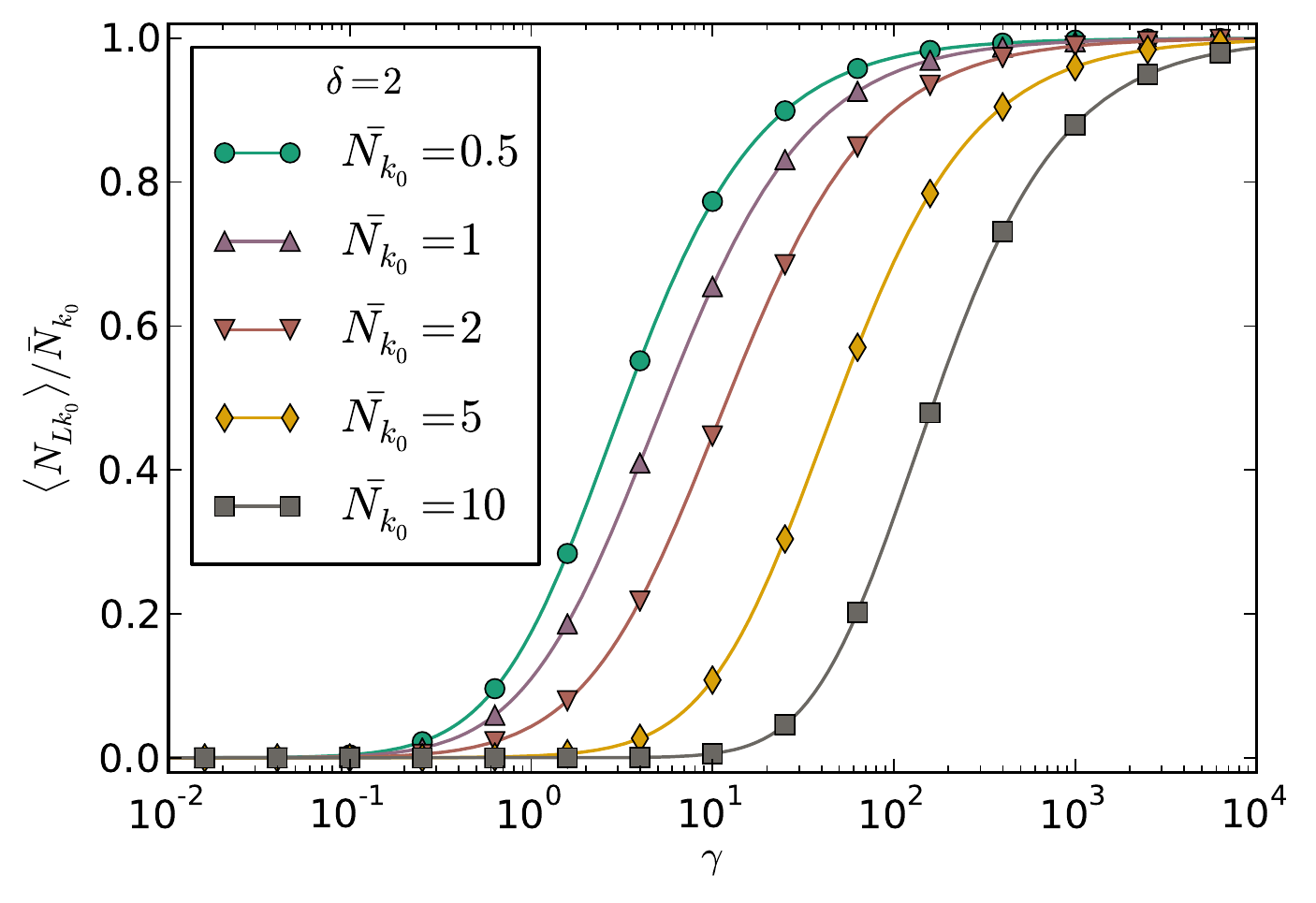}
\includegraphics[width=0.48\textwidth]{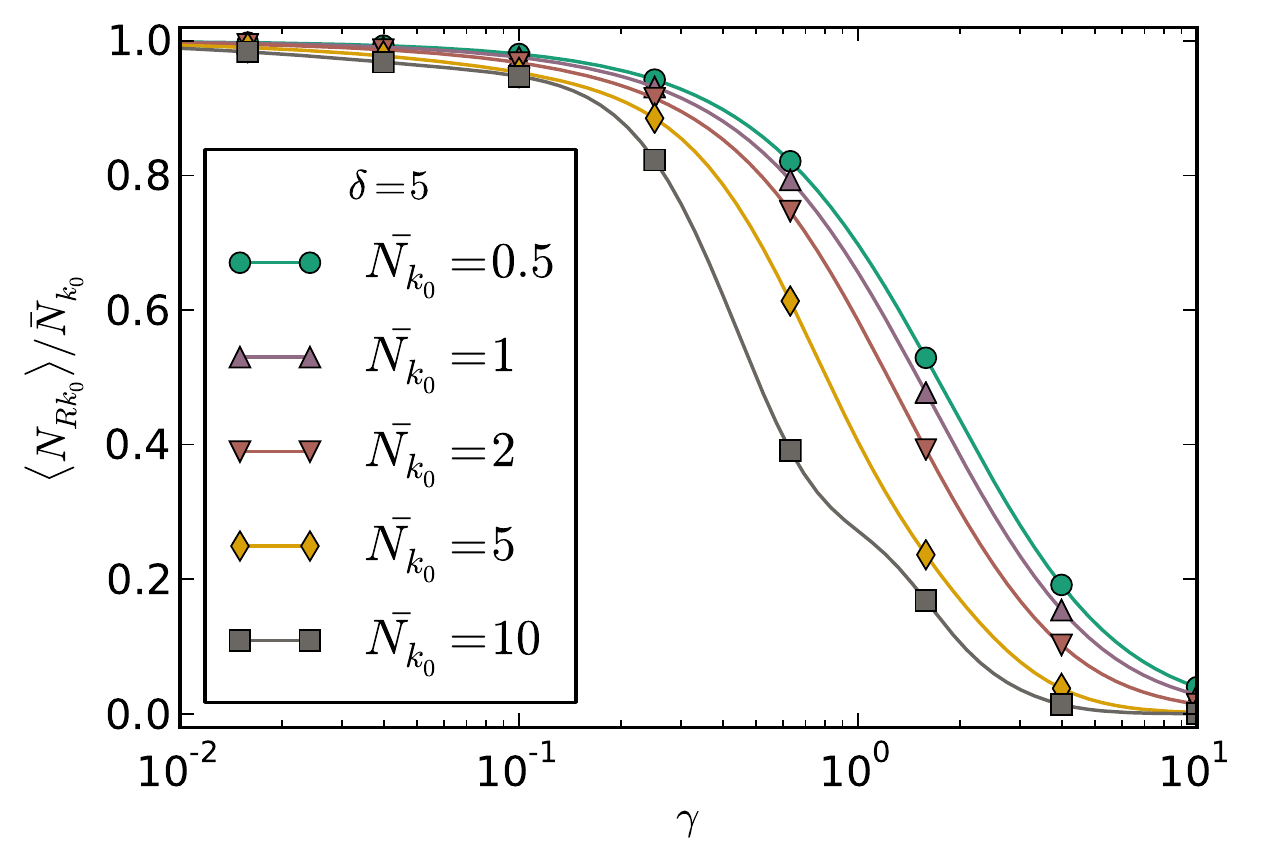}
\includegraphics[width=0.48\textwidth]{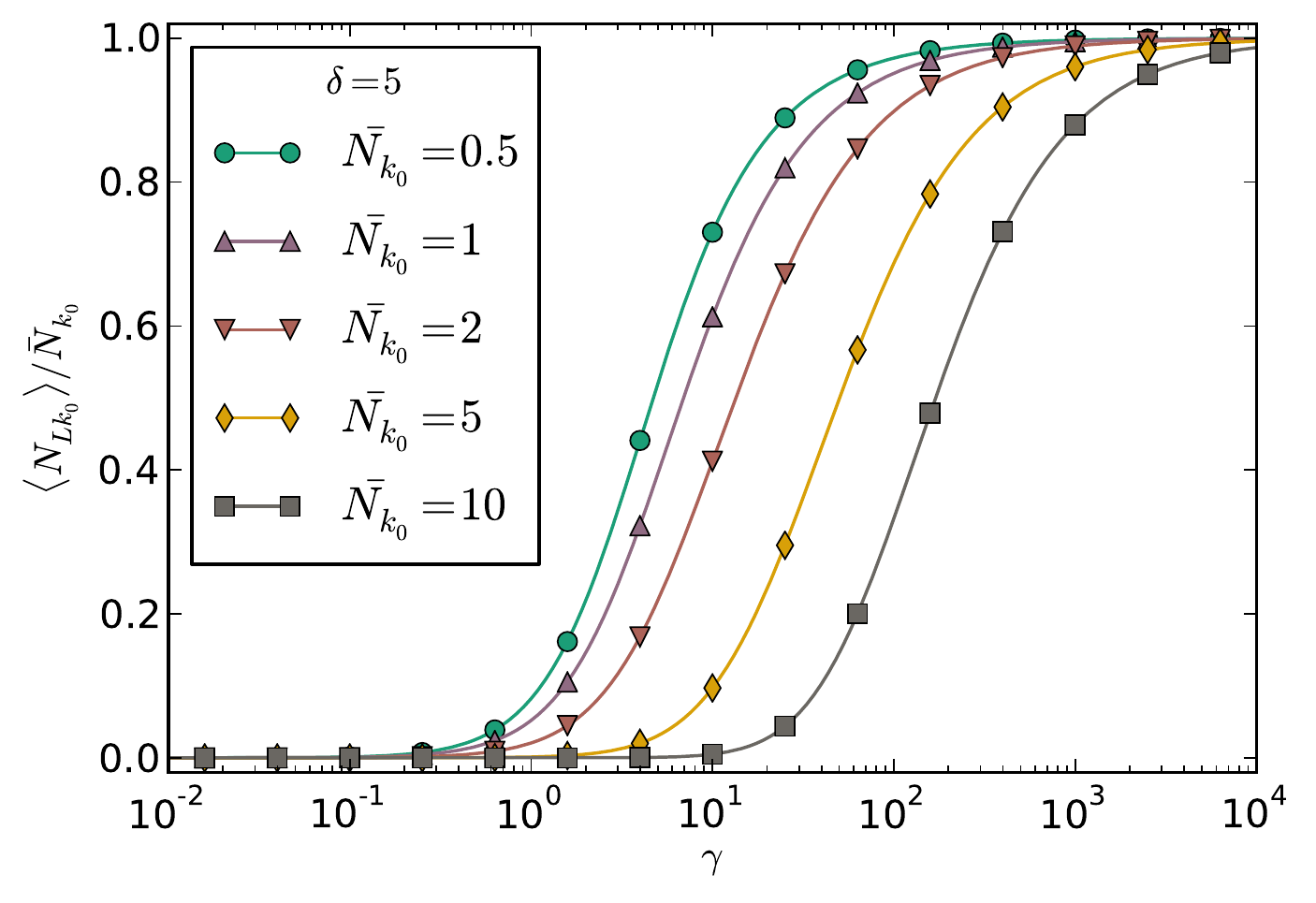}
    \caption{Evolution of the mean number of particles remaining in the mode $k_0$ in right and left channels as a function of the coupling strength, initial population, and detuning. }
    \label{fig:rl-mean}
\end{figure}
\begin{figure}[t]
       \includegraphics[width=0.48\textwidth]{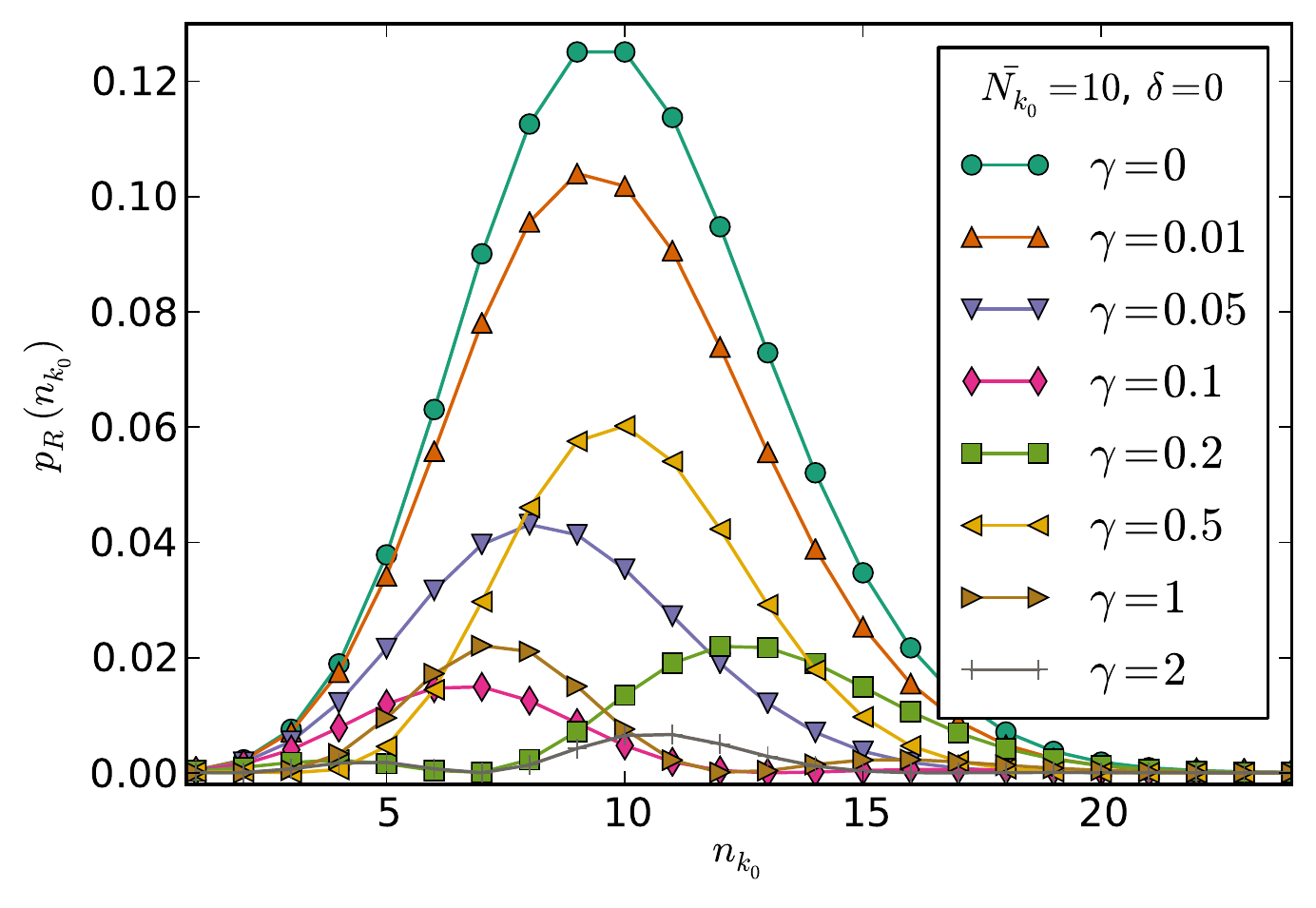}
       \includegraphics[width=0.48\textwidth]{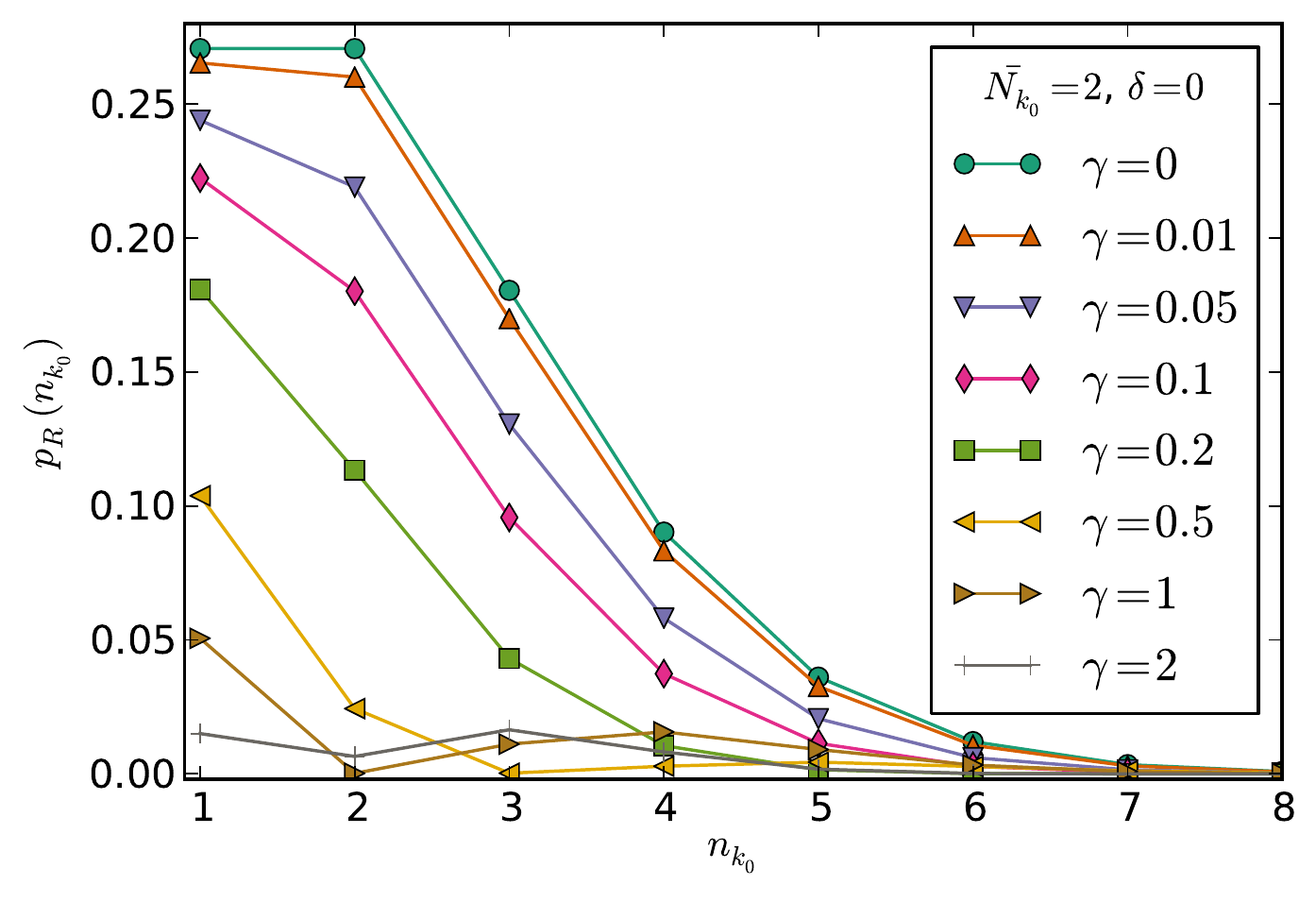}
    \caption{The probability
    distribution~$p_{r}(n)$ for the number of particles in the mode~$k_0$ in the $r$-channel, plotted for various values of~$\bar{N}$ and the interaction
    strength~$\gamma$. 
    }
\end{figure}
\begin{figure}[t]
       \includegraphics[width=0.48\textwidth]{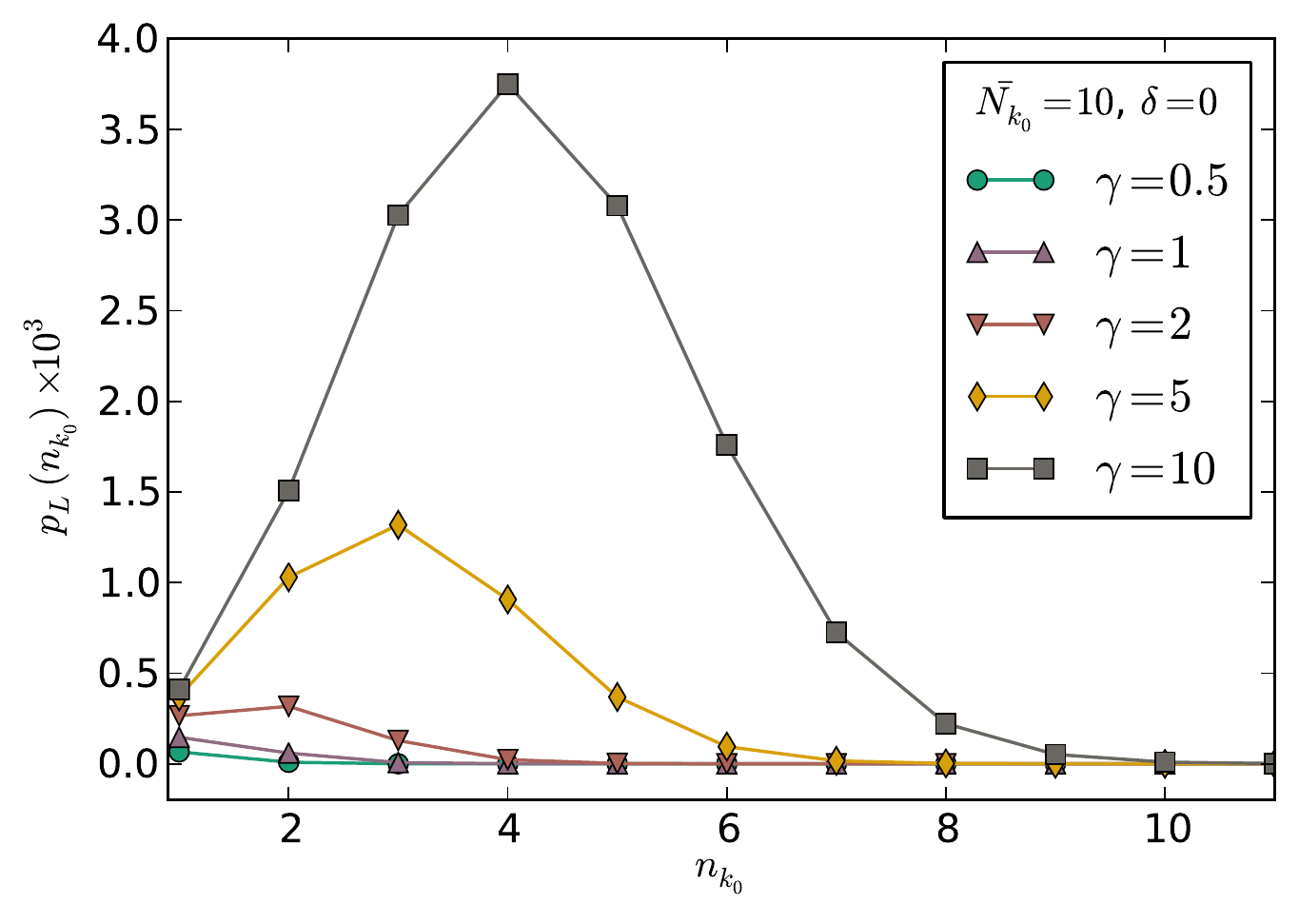}
       \includegraphics[width=0.48\textwidth]{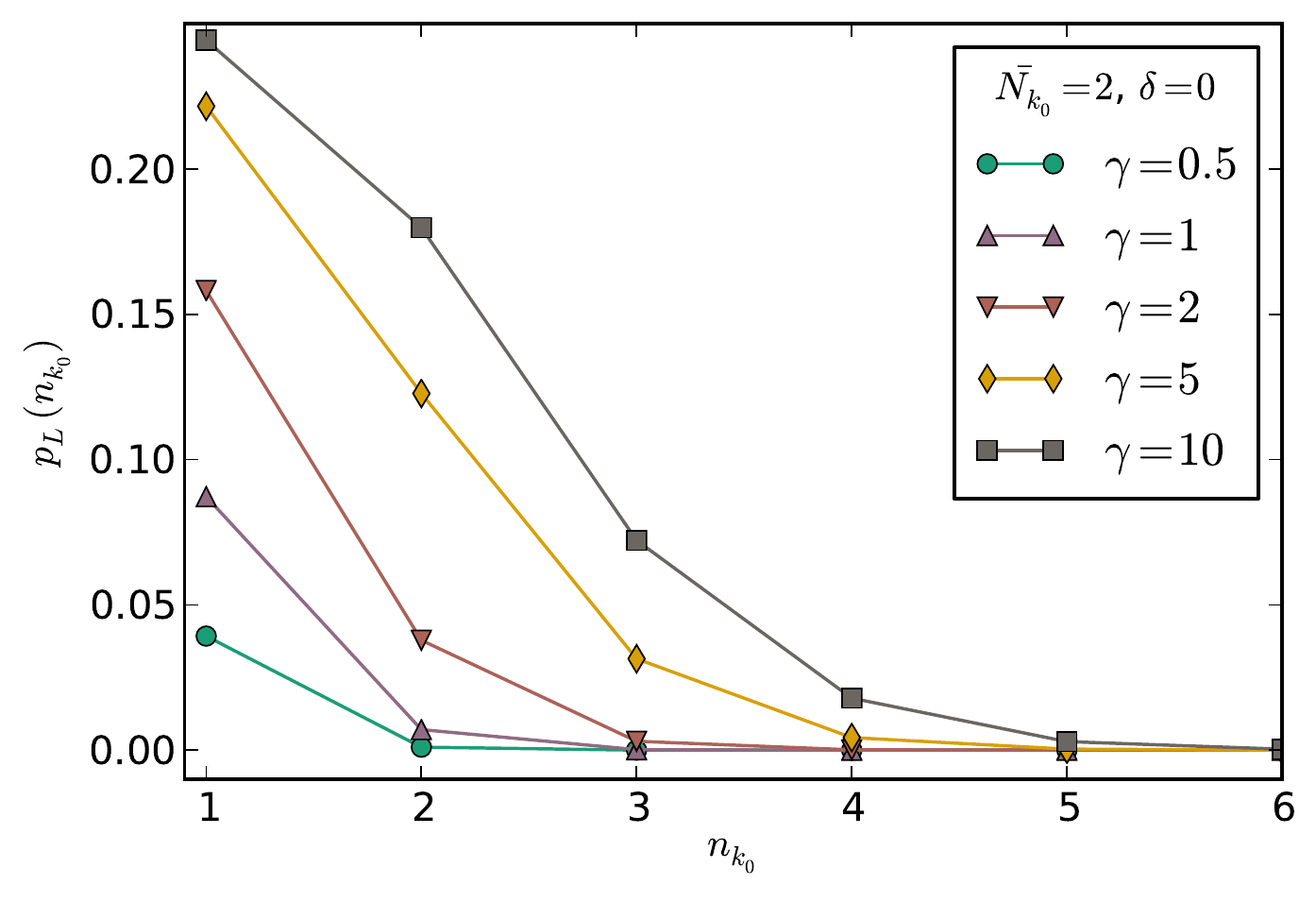}
    \caption{The probability
    distribution~$p_{l}(n)$ for the number of particles in the mode~$k_0$ in the $l$-channel, plotted for various values of~$\bar{N}$ and the interaction
    strength~$\gamma$. 
    }
\end{figure}


\begin{thebibliography}{99}

\bibitem{Glauber}
R. J. Glauber, Quantum theory of optical coherence, Wiley-VCH, 2007.

\bibitem{Mandel}
L. Mandel, Opt. Lett. {\bf 4}, 205 (1979).

\bibitem{KK}
P. L. Kelley and W. H. Kleiner, Phys. Rev. {\bf 136}, A316 (1964).

\bibitem{LL}
L. S. Levitov, G. B. Lesovik, JETP Lett. {\bf 58}, 230 (1993); L. S. Levitov, H. Lee, G. B. Lesovik, J. Math. Phys. {\bf 37}, 4845 (1996).

\bibitem{NB}
Y. V. Nazarov, Ya. M. Blanter, Quantum Transport: Introduction to Nanoscience (Cambridge UP, 2009); W. Belzig, arXiv:cond-mat/0312180.



\bibitem{BN}
D. A. Bagrets, Yu. V. Nazarov, Phys. Rev. B {\bf 67}, 085316 (2003); D. A. Bagrets, Y.  Utsumi, D. S. Golubev, and G. Sch\"{o}n, Fortschr. Phys. {\bf 54}, 917 (2006).


\bibitem{GK}
A. O. Gogolin and A. Komnik, Phys. Rev. Lett. {\bf 97}, 016602 (2006).

\bibitem{Schonh}
K. Sch\"{o}nhammer, Phys. Rev. B {\bf 75}, 205329 (2007).



\bibitem{KS}
A. Komnik, H. Saleur, Phys. Rev. Lett. {\bf 107}, 100601 (2011).

\bibitem{BSS}
E. Boulat, H. Saleur, P. Schmitteckert, Phys. Rev. Lett. {\bf 101}, 140601 (2008); 
A. Bransch\"{a}del, E. Boulat, H. Saleur, P. Schmitteckert, Phys. Rev. Lett. {\bf 105}, 146805 (2010).


\bibitem{CBS}
S. T. Carr, D. A. Bagrets, P. Schmitteckert, Phys. Rev. Lett. {\bf 107}, 206801 (2011).




\bibitem{SB}
E. V. Sukhorukov and O. M. Bulashenko, Phys. Rev. Lett. {\bf 94}, 116803 (2005).

\bibitem{KL}
I. Klich and L. Levitov, Phys. Rev. Lett. {\bf 102}, 100502 (2009).

\bibitem{IA}
D. A. Ivanov and A. G. Abanov, Europhys. Lett. {\bf 92}, 37008 (2010).


\bibitem{GADP}
V. Gritsev et al, Nature Phys. {\bf 2}, 705 (2006); A. Imambekov et al, cond-mat/0703766.

\bibitem{Hoffer}
S. Hoffereberth et al, Nature Phys. {\bf 4}, 489 (2008).

\bibitem{LF}
A. Lamacraft and P. Fendley, Phys. Rev. Lett. {\bf 100}, 165706 (2008).

\bibitem{prethermalization}
M. Gring et al, Science {\bf 337}, 1318 (2012). 

\bibitem{IA2}
D. A. Ivanov and A. G. Abanov, arXiv:1203.6325.



\bibitem{QOnano}
A. V. Akimov et al., Nature {\bf 450}, 402 (2007); O. Astafiev et al, Science {\bf 327}, 840 (2010); 
B. Dayan et al., Science {\bf 319}, 1062 (2008).







\bibitem{few-photons}
K. Kojima, H. F. Hofmann, S. Takeuchi, K. Sasaki, Phys. Rev. A {\bf 68}, 013803 (2003); J.-T. Shen, S, Fan, Phys. Rev. Lett. {\bf 95}, 213001 (2005); 
Phys. Rev. Lett. {\bf 98}, 153003 (2007);
 D. E. Chang, A. S. Sorensen, E. A. Demler, M. D. Lukin, Nature Phys. {\bf 3}, 807 (2007);
L. Zhou, Z. R. Gong, Y. Liu, C. P. Sun, F. Nori, Phys. Rev. Lett. {\bf 101}, 100501 (2008); P. Longo, P. Schmitteckert, K. Busch, Phys. Rev. Lett. {\bf 104}, 023602 (2010); D. Witthaut , A. S. Sorensen, New J. Phys. {\bf 12}, 043052 (2010);
D. Roy, Phys. Rev. Lett. {\bf 106}, 053601 (2011); 
H. Zheng, D. J. Gauthier, H. U. Baranger 
Phys. Rev. A {\bf 85}, 043832 (2012). 


\bibitem{supplement}
In the Supplementary Material we outline essential technical details of our derivations and present some additional results.

\bibitem{PGnjp}
M. Pletyukhov and V. Gritsev, New J. Phys.
{\bf 14}, 095028 (2012).

\bibitem{Mussardo}
G. Mussardo, Statistical Field theory (Oxford UP, 2010).




\bibitem{Mollow}
B. R. Mollow, Phys. Rev. {\bf 188}, 1969 (1969). 



\bibitem{earlier}
R. J. Cook, Phys. Rev. A {\bf 23}, 1243 (1981); D. Lenstra, Phys. Rev. A {\bf 26}, 3369 (1982).

\bibitem{BB}
G. Bel, F. Brown, Phys. Rev. Lett. {\bf 102}, 018303 (2009).


\bibitem{PRG-long}
M. Pletyukhov, M. Ringel, and V. Gritsev, in preparation.






\bibitem{Eichler}
C. Eichler et al, 
Phys. Rev. Lett. {\bf 109}, 240501 (2012); C. Eichler et al, 
Phys. Rev. Lett. {\bf 106}, 220503 (2011).




\end{thebibliography}
\end{document}